\documentclass[aip,jcp,preprint]{revtex4-1}

\usepackage[T1]{fontenc}       
\usepackage{graphicx}
\usepackage{dcolumn}
\usepackage{bm}
\usepackage{amsmath}
\usepackage{amsfonts}
\usepackage{amssymb}
\usepackage{braket}
\usepackage[separate-uncertainty = true,multi-part-units=single]{siunitx}
\usepackage{subcaption}
\usepackage{color}
\usepackage{hyperref}
\usepackage{epstopdf}
\AppendGraphicsExtensions{.tif}

\let\oldsqrt\sqrt
\def\sqrt{\mathpalette\DHLhksqrt}
\def\DHLhksqrt#1#2{%
\setbox0=\hbox{$#1\oldsqrt{#2\,}$}\dimen0=\ht0
\advance\dimen0-0.2\ht0
\setbox2=\hbox{\vrule height\ht0 depth -\dimen0}%
{\box0\lower0.4pt\box2}}

\begin{document}
\preprint{A18.04.0117}

\title{Dynamical structure of entangled polymers simulated under shear flow}

\author{Airidas Korolkovas}
\affiliation{Institut Laue-Langevin, 71 rue des Martyrs, 38000 Grenoble, France}
\affiliation{Division for Material Physics, Department for Physics and Astronomy, Lägerhyddsvägen 1, 752 37 Uppsala, Sweden}
\email{korolkovas@ill.fr}

\author{Philipp Gutfreund}
\affiliation{Institut Laue-Langevin, 71 rue des Martyrs, 38000 Grenoble, France}

\author{Max Wolff}
\affiliation{Division for Material Physics, Department for Physics and Astronomy, Lägerhyddsvägen 1, 752 37 Uppsala, Sweden}

\date{\today}

\begin{abstract}
The non-linear response of entangled polymers to shear flow is complicated. Its current understanding is framed mainly as a rheological description in terms of the complex viscosity. However, the full picture requires an assessment of the dynamical structure of individual polymer chains which give rise to the macroscopic observables. Here we shed new light on this problem, using a computer simulation based on a blob model, extended to describe shear flow in polymer melts and semi-dilute solutions. We examine the diffusion and the intermediate scattering spectra during a steady shear flow. The relaxation dynamics are found to speed up along the flow direction, but slow down along the shear gradient direction. The third axis, vorticity, shows a slowdown at the short scale of a tube, but reaches a net speedup at the large scale of the chain radius of gyration.
\end{abstract}

\maketitle

\section{Introduction}
\begin{figure}[ptbh!]
\begingroup
			\sbox0{\includegraphics{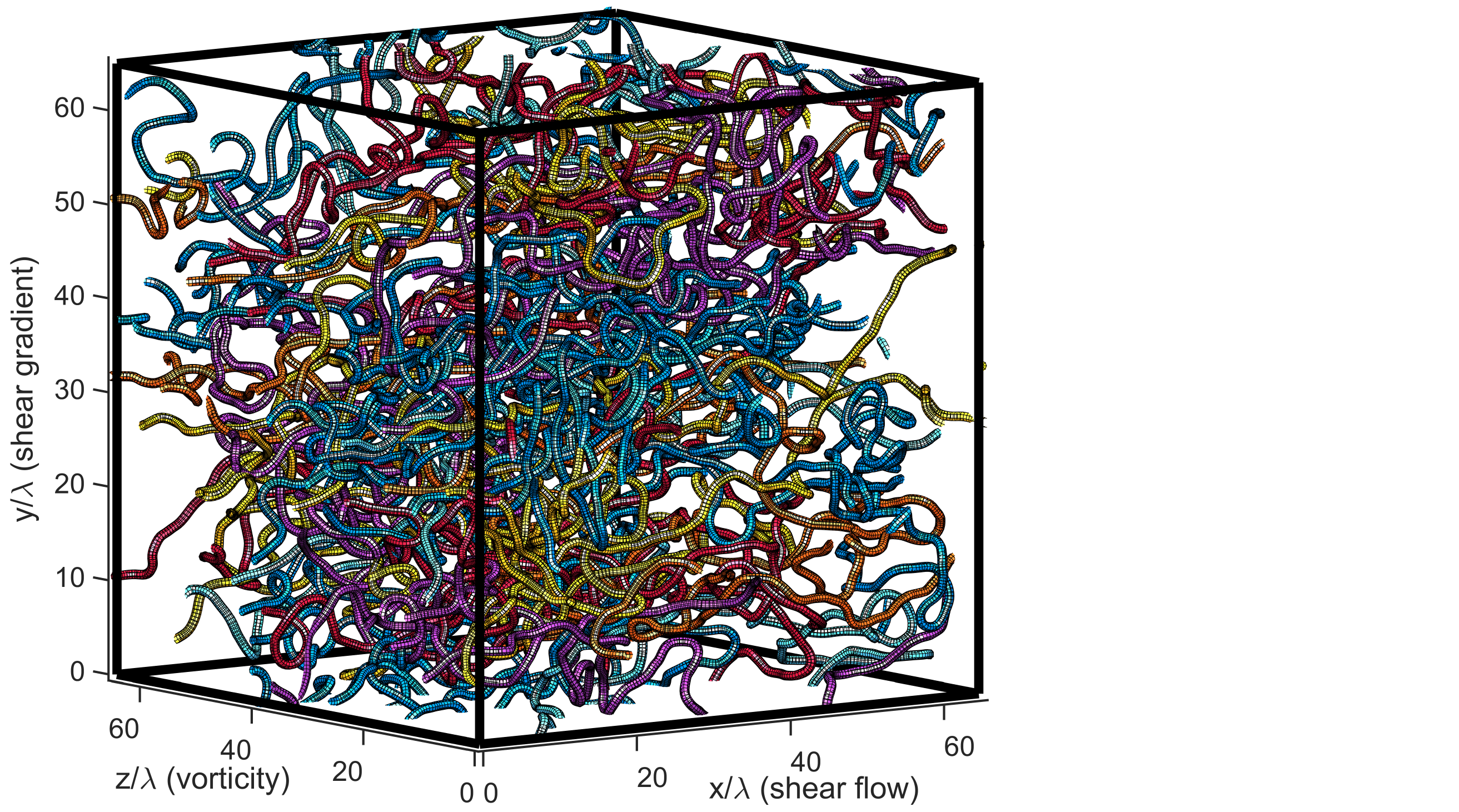}} 
			\includegraphics[clip,trim={.0\wd0} 0 {.2\wd0} 0,width=0.6\linewidth]{./figures/3Dshear.pdf} 
		\endgroup
\caption{A simulation box with $C=64$ chains under shear rate $\text{Wi}=32$. Every chain has $N=512$ blobs, or degrees of freedom. The radius of the blob defines the length scale $\lambda$. For visual clarity, the diameter of the plotted tubes is set to $\lambda$ as well. Each molecule is colored in one of seven randomly chosen hues.}\label{3Dconformation}
\end{figure}

Entangled polymers are a network of long, interpenetrating chains, as illustrated in Fig.~\ref{3Dconformation}. The motion is driven by random thermal fluctuations, while the excluded volume repulsion prevents the chains from crossing each other. The resulting dynamics display a rich variety of phenomena, distinct from most other materials. Despite its macroscopically complicated behaviour, the physics of the polymer network is mostly governed by one parameter: the number of entanglements $Z=N/N_e$, which is proportional to the chain length $N$. The entanglement length $N_e$ determines the onset of multi-chain effects, and its value may depend on the chemical composition~\cite{cho2004universality}, temperature~\cite{cavaille1987time}, and concentration~\cite{kasaai2000master, wingstrand2015linear}. However, these parameters typically do not influence the macroscopic properties, but only shift the time and the length scales. This universality suggests that most of polymer physics might be described in terms of one fundamental model, perhaps containing only one parameter, $Z$.

Theoretical descriptions usually simplify the full multi-chain network with just one chain immersed in the averaged confining field of all the other chains. Currently the main reference is the GLaMM theory~\cite{graham2003microscopic}, rooted in the earlier Doi and Edwards theory~\cite{doi1988theory}, itself based on de Gennes work~\cite{de1971reptation}. It is widely applied in rheology~\cite{han2007rheology} where a strong shear flow results in a drop of viscosity (shear thinning), and the emergence of normal stress differences (the Weissenberg effect). The theory also corroborates structural measurements, in particular small angle neutron scattering (SANS) and neutron reflectometry (NR). These experiments have established that the polymer shape elongates in the shear flow direction $x$, while shrinking perpendicular to it~\cite{chenneviere2016direct, korolkovas2017polymer,Yearley2010, Bent2003, Graham2006, Clarke2010, Muller1993, Blanchard2005, Boue1987, Kirkensgaard2016}, along both shear gradient $y$ and vorticity $z$. However, there is some disagreement with the theory on the exact relaxation pathways as evidenced by a recent kinetic SANS experiment~\cite{wang2017fingerprinting}.

Rheology and SANS provide a fundamentally static molecular picture: the stress tensor and the average distance between labeled monomers. The time evolution of these quantities can also be measured in kinetic experiments~\cite{wang2017fingerprinting}. The next level of understanding is the dynamical structure, which tells us how fast the various segments of the polymer are fluctuating and how correlated are they. While numerous dynamical experiments have been published~\cite{goossen2015sensing, schleger1998clear, wischnewski2002molecular, richter1993dynamics, richter1993origins, pyckhout2016mixtures}, they are mostly limited to equilibrium fluctuations. By contrast, practical and industrial use of entangled polymers usually involves flow, with examples in food thickeners~\cite{rao2014rheology}, paint formulations~\cite{chhabra2011non}, plastics injection~\cite{bartczak2005effect} and extrusion~\cite{bartczak2010effect}, fibre spinning~\cite{shenoy2005role}, and enhanced oil recovery from shale rock~\cite{wever2011polymers}. Non-equilibrium entanglement dynamics also play a biological role in areas like the cytoskeleton~\cite{tsang2017dynamic} during cell division, and the cartilage~\cite{june2009cartilage} during joint movement.

The most complicated, but potentially very rewarding experiments aim to measure polymer dynamics while in the non-equilibrium state. Such information quantifies the molecular motion directly, rather than having to infer it from the change of static structure, as in kinetic SANS. However, due to practical limitations, dynamical experiments are rare, although not impossible. One example is video microscopy of entangled DNA, which can record the shape fluctuations of an individual stained chain as it undergoes flow~\cite{teixeira2007individualistic}. A very broad range of chain conformations were detected, raising doubts about the validity of the mean-field assumption used in tube theories. Another option is superposition rheology (SR), where a small oscillatory force is applied on top of a steady shear flow, probing the mechanical relaxation as it is altered by shear. Using solutions of entangled polyisobutene (PIB), the authors have observed a speedup of the relaxation spectrum, and explained it with a macroscopic Wagner model~\cite{vermant1998orthogonal}. In the case of entangled polyisoprene (PI) solutions, a slight speedup along the shear gradient axis was reported using dielectric spectroscopy (DS)~\cite{watanabe2003rheo}, which probes the motion of the whole molecule. The measured effect is much smaller than the tube theory prediction. By contrast, a clear slowdown has been observed with nuclear magnetic resonance (NMR) spectroscopy~\cite{bohme2011nmr} probing internal motions of an entangled polypropylene (PP) solution. Recently, neutron spin echo (NSE) was attempted under shear flow, reporting no change along the vorticity axis within the accessible shear rates using an entangled polydimethylsiloxane (PDMS) melt~\cite{Kawecki2017}.

Perhaps due to the scarcity of reliable experiments, theoretical work has largely neglected the question of non-equilibrium dynamics. Multi-chain computer simulations is a useful addition, since they do not involve any mean field assumptions and can verify, refute, or complement the effective single chain theories. However, just like in experiments, most published simulations have focused on the average steady-state~\cite{aoyagi2000molecular, kroger2000rheological, padding2003coarse, baig2010flow, dorgan2012parameter, halverson2012rheology, xu2014shear} and occasionally the transient response~\cite{kroger1997polymer, cao2015simulating, xu2018molecular}. More recent simulations have started to probe the non-equilibrium relaxation dynamics, for both short~\cite{kim2011mean, huang2012non} and long entangled chains~\cite{nafar2015individual, mohagheghi2016elucidating, sefiddashti2017evaluation}. Some disagreements with tube theories have been pointed out, questioning their validity under strong shear, especially as far as dynamics is concerned.

Clearly, more experiments are needed to resolve the controversies. In this paper we provide the multi-chain simulation data for comparison with anticipated future experiments. We show the specific 3D pathway of how the diffusion morphs from isotropic at short time scales, to anisotropic at the longest time scales. An intriguing behaviour is found along the vorticity axis, where the change of dynamics is non-monotonic: it slows down on the short scale, but speeds up on the long scale, helping to reconcile experimental evidence. Our results will encourage new experiments and contribute to further developments in tube theory.

\section{Simulation method}
The simulation is based on a previously published algorithm~\cite{korolkovas2016simulation} for equilibrium systems, but now extended for bulk shear flow by imposing a Couette velocity profile, described in detail in Appendix A. The code in MATLAB and CUDA is available for download from a Zenodo repository~\cite{airidas_korolkovas_2017_439529}. We numerically solve the motion of $c=1,\, 2,\, \ldots,\, C$ chains in a box, each described by a continuous curve $\mathbf{R}_c(t,s)$, with variables $t$ for time and $0<s<1$ for monomer index. There are $N$ degrees of freedom per chain, governed by the first order equation of motion:
\begin{subequations}\label{eqmotion}
\begin{align}
\zeta \frac{\partial \mathbf{R}_c (t,s)}{\partial t} &= \left(\frac{3k_B T}{Nb^2}\right) \frac{\partial^2 \mathbf{R}_c(t,s)}{\partial s^2} &&  \text{(spring)}\\
&+ \frac{Nv}{\lambda^3} \sum_{c'=1}^C \int_0^1 ds'\,  F[\mathbf{R}_c(t,s) - \mathbf{R}_{c'}(t,s')] && \text{(excluded volume)}\\
&+ \zeta \dot{\gamma} \left( \mathbf{R}_c(t,s) \cdot \mathbf{e}_y \right) \mathbf{e}_x  && \text{(shear flow)}\\
&+ \sqrt{2k_B T \zeta} \mathbf{W}_c (t,s) && \text{(thermal noise)}
\end{align}
\end{subequations}
It describes the motion of a continuous Rouse chain (Eq.~(4.9) in Ref.~\cite{doi1988theory}), with an added excluded volume field that we have implemented in Ref.~\cite{korolkovas2016simulation}, as well as a shear field that is described in Appendix A of this article. Here $\zeta = 6\pi \eta_s b N$ is the friction coefficient of the chain center of mass, defining the unit of time:
\begin{equation}\label{tau}
\tau = \frac{6\pi \eta_s b^3}{k_B T},
\end{equation}
where $b$ is the spring length. The thermal noise is modeled by a Wiener process $\braket{\mathbf{W}_c(t,s)\mathbf{W}_{c'}(t',s')} = \mathbb{I}\, \delta_{cc'}\,\delta(t-t')\,\delta(s-s')$. The repulsive excluded volume force between the chains is described by the blob model, which is well approximated~\cite{bolhuis2001accurate} by a Gaussian function:
\begin{equation}\label{forcefield}
F(\mathbf{r}) = \frac{k_B T\mathbf{r}}{\lambda^2} e^{-\mathbf{r}^2/2\lambda^2}.
\end{equation}
For a semi-dilute polymer solution~\cite{deGennesScalingConcepts}, the blob radius $\lambda$ and the number of blobs per chain $N$ scale as
\begin{equation}\label{mapping}
\lambda \propto \rho^{-3/4} \quad \text{and} \quad N \propto \rho^{5/4} M_w
\end{equation}
where $\rho$ is the density and $M_w$ is the molecular weight. The solvent is implicitly contained in the blob particle and its effect is taken into account via the viscosity $\eta_s$ (Eq.~\eqref{tau}). This model can also be extended for polymer melts, $\rho \rightarrow \rho_0$, although in that case the mapping of Eq.~\eqref{mapping} does not apply and the absolute values of $\lambda$ and $N$ depend on the chemical species and temperature. 

Entire studies have been devoted to relate the coarse interaction parameters with the atomistic details of a particular chemical species~\cite{sun2005systematic}. However, within the numerical pre-factor, the physical properties on the large scale are found to be universal across vastly different polymer models, ranging from lattice-based~\cite{dorgan2012parameter}, to hard bead-and-spring~\cite{kremer1990dynamics}, to our soft blobs~\cite{korolkovas2016simulation}. We do not attempt a one-to-one correspondence with any particular experiment, and for a generic model we pick the natural choice of the excluded volume $v = \lambda^3 = b^3$. With these settings, the excluded volume and the spring forces have the same strength, enabling a maximum time step. Other simulations often use a bending potential~\cite{cao2012shear}, which has the advantage of decreasing the entanglement length $N_e$. In our case, we could obtain a similar effect by increasing the chain stiffness $\lambda/b > 1$, but that would also require increasing $v$ to prevent the contour from shrinking, and this reduces the time step since $\Delta t \propto 1/v$, thus offsetting the benefit of a smaller $N_e$.

While the chain crossings are not strictly forbidden, their probability can be suppressed sufficiently so that highly entangled dynamics emerge, as explained in our equilibrium publication~\cite{korolkovas2016simulation}. First, the continuous equation of motion~\eqref{eqmotion} is sampled with $J$ discrete points, and choosing $J\gg N$ guarantees that the gaps between those points remain much smaller than the blob radius. In other words, there are more interaction centers ($J$) than degrees of freedom ($N$), enabling the time step $\Delta t$ to be on the large scale of the $\lambda$-sized blobs $N$, rather than on the much shorter scale of the $J$-points. A special case of $J=N$ would correspond to standard bead and spring simulations. Second, the instantaneous random force must be much weaker than the excluded volume force. Note that we are working with a first order equation of motion, which is overdamped and does not conserve momentum, unlike the second order equation of motion. In this case, we can decrease the random force magnitude by $1/\sqrt{M}$, provided that we only update its direction once per $M$ steps, effectively cutting off high frequency fluctuations, while maintaining the mean square displacement fixed in the long run. If the cutoff time $M\Delta t$ remains much smaller than the entanglement time $\tau_e$ (Eq.~\eqref{taue}), the physics on the long time scale are not affected, and a value of $M=120$ was found to be a suitable compromise. A similar argument is made in the modified Kremer-Grest models, where a bending potential is introduced, justified by the fact that it does not influence the physics, as long as the chain persistence length remains much shorter than the entanglement length~\cite{bacova2013dynamics}.

When shear is applied, the probability of chain crossings obviously increases. However, the increase remains negligible for shear rates below the inverse Rouse time: $\dot{\gamma} \tau N^2 \lesssim 1$, where the main effect is the reorientation of the entanglement network. We do not apply a higher shear, since the chain backbone begins to stretch at that point. We use chains of length $N=512$, shorter than the longest one in our equilibrium study. The appeal of working with a coarse model is that high frequency fluctuations are not interesting on the large scale~\cite{ilg2009systematic}, which automatically precludes an excessively strong shear. Rheological experiments seldom go beyond $\text{Wi} = 10$, and there is plenty of physics to be learned even in a modestly non-Newtonian regime. To the contrary, the ultra high shear rates of order $\SI{e8}{\second^{-1}}$, routinely considered in atomistic polymer simulations~\cite{kroger2000rheological}, are never realized experimentally, and may only occur during a cataclysmic event like an asteroid impact~\cite{ramesh2008}. Moreover, strong shear is known to destroy polymers by rupturing covalent bonds~\cite{nghe2010flow}.

In general, computer simulations may suffer from a number of technical biases: the chains are too stiff, the box is too small, the system is not properly equilibrated, and there may be chain crossings. These biases all result in faster apparent dynamics, while trying to minimize them makes the program run slower, limiting the accessible length and time scales. We have strived to reach a reasonable compromise between these limitations. Dedicated studies are necessary to fully assess chain crossings~\cite{padding2001uncrossability}, going beyond the scope of the present article. In the meanwhile, we caution that our results may indeed have a slight bias, more so at the highest shear rates and the longest time scales.

\section{Results}
The simulation was used to explore the changes in relaxation dynamics of polymers when subject to a steady shear flow. We have simulated $C=64$ chains of length $N=512$ in a cubic box of volume
\begin{equation}
V = L^3 = v_0\left(\frac{4\pi}{3}\right)NC\lambda^3
\end{equation}
with a numerical pre-factor $v_0 = 2$ to allow some freedom of movement for the blobs. The microscopic parameters are set to $v=\lambda^3=b^3$. This melt of blobs is moderately entangled, with about $Z = N/N_e = 8.5$ entanglements per chain at equilibrium, as determined topologically using the Z1 code~\cite{karayiannis2009combined}. Seven simulation runs where performed with all parameters kept fixed except for the shear rate $\text{Wi} = \dot{\gamma}\tau_d = 0,\, 0.3,\, 1,\, 3,\, 10,\, 32,\, \text{and}\, 100$. The computations were carried out on an Nvidia Quadro P5000 GPU card for about 4 days per run, generating trajectories of $20\tau_d$ relaxation times each. To avoid any start-up effect in the subsequent data analysis, we have discarded the first $\tau_d$ steps of each trajectory, so the data reported in this article is for a steady-state shear.

\begin{figure}[ptbh!] 
\begin{subfigure}{.49\textwidth}
		\begingroup
			\sbox0{\includegraphics{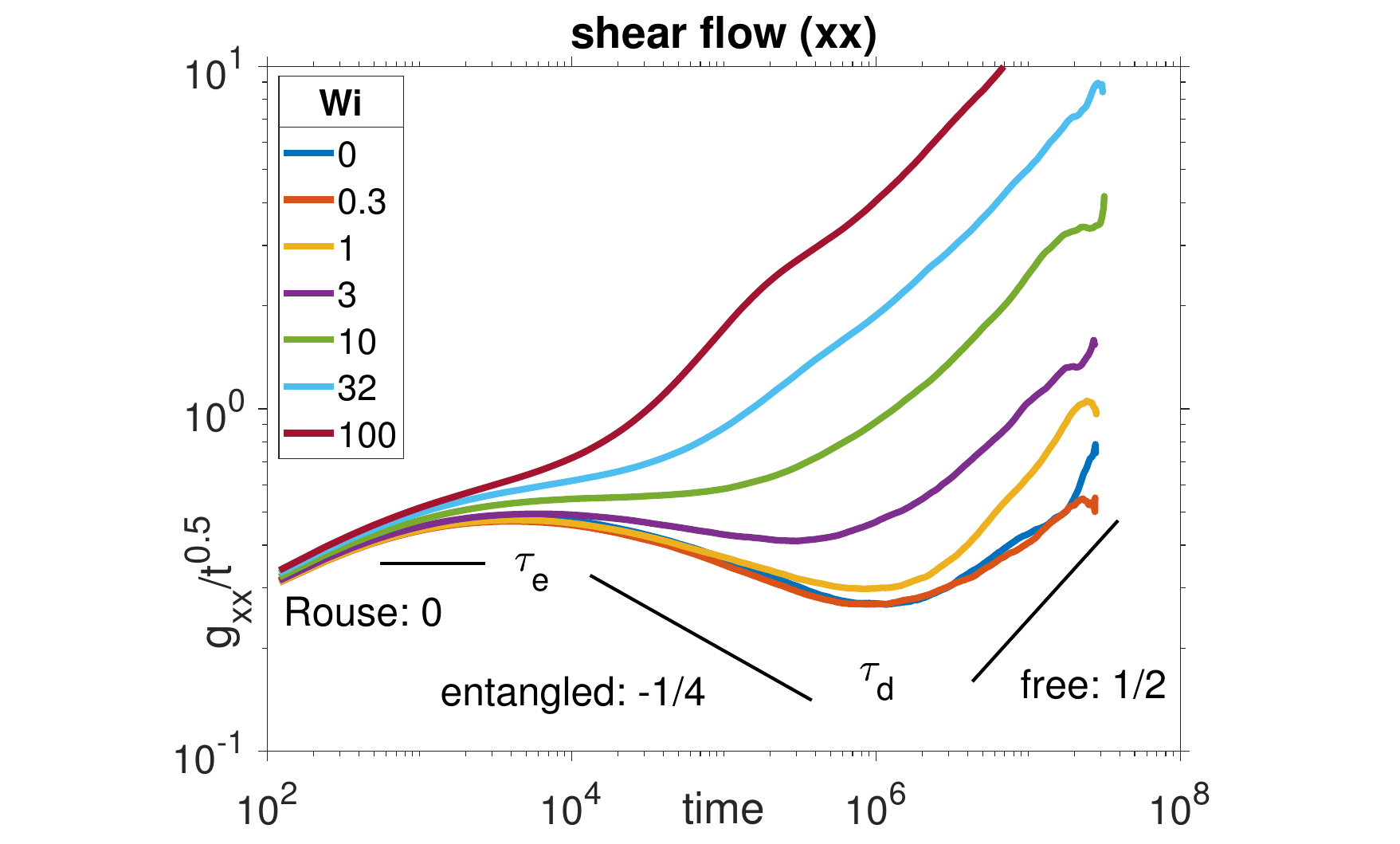}} 
			\includegraphics[clip,trim={.1\wd0} 0 {.13\wd0} 0,width=\linewidth]{./figures/gxx.pdf} 
		\endgroup
\end{subfigure}
\hfill
\begin{subfigure}{.49\textwidth}
\begingroup
			\sbox0{\includegraphics{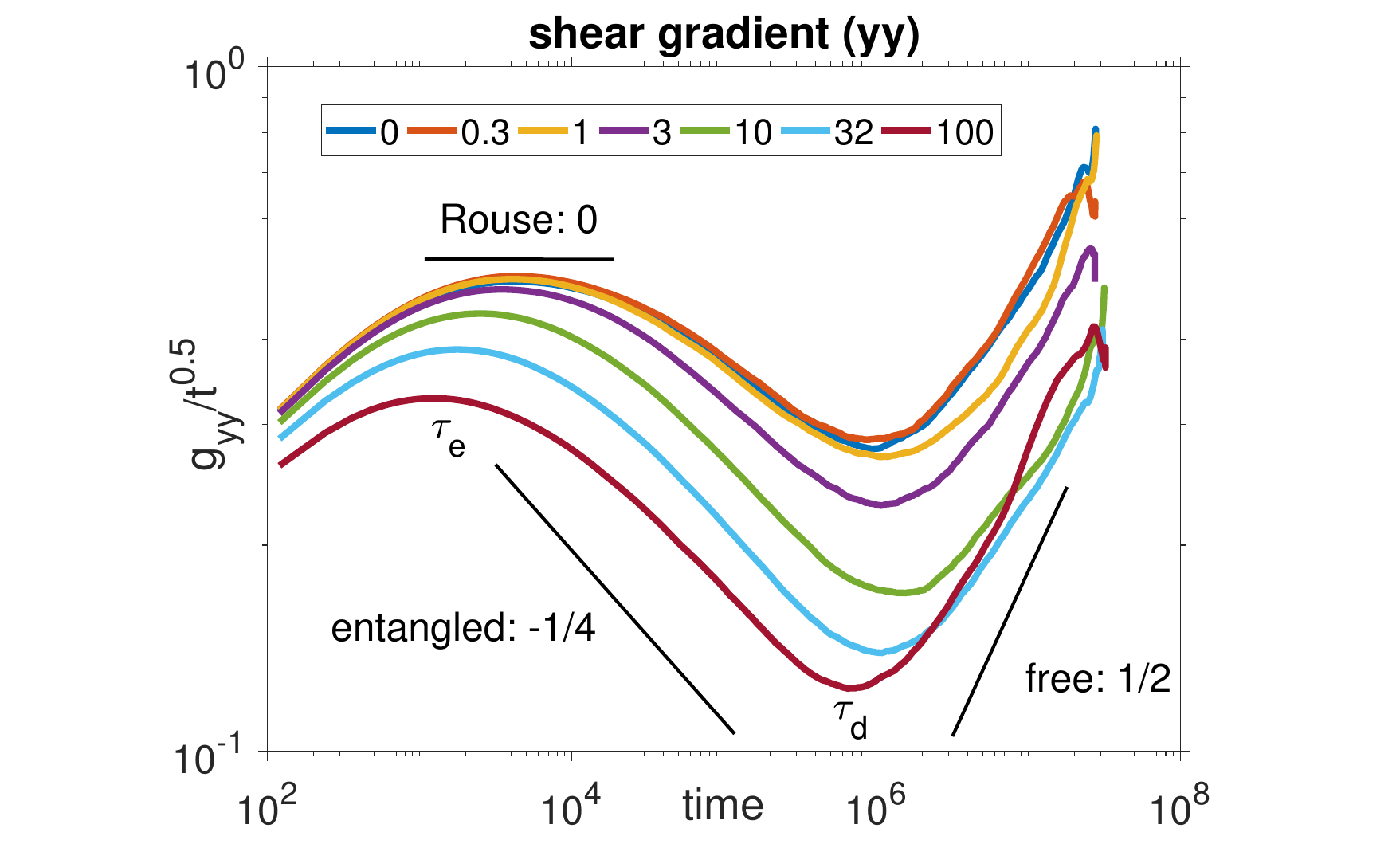}} 
			\includegraphics[clip,trim={.1\wd0} 0 {.13\wd0} 0,width=\linewidth]{./figures/gyy.pdf} 
		\endgroup
\end{subfigure}
\vskip\baselineskip
\begin{subfigure}{.49\textwidth}
\begingroup
			\sbox0{\includegraphics{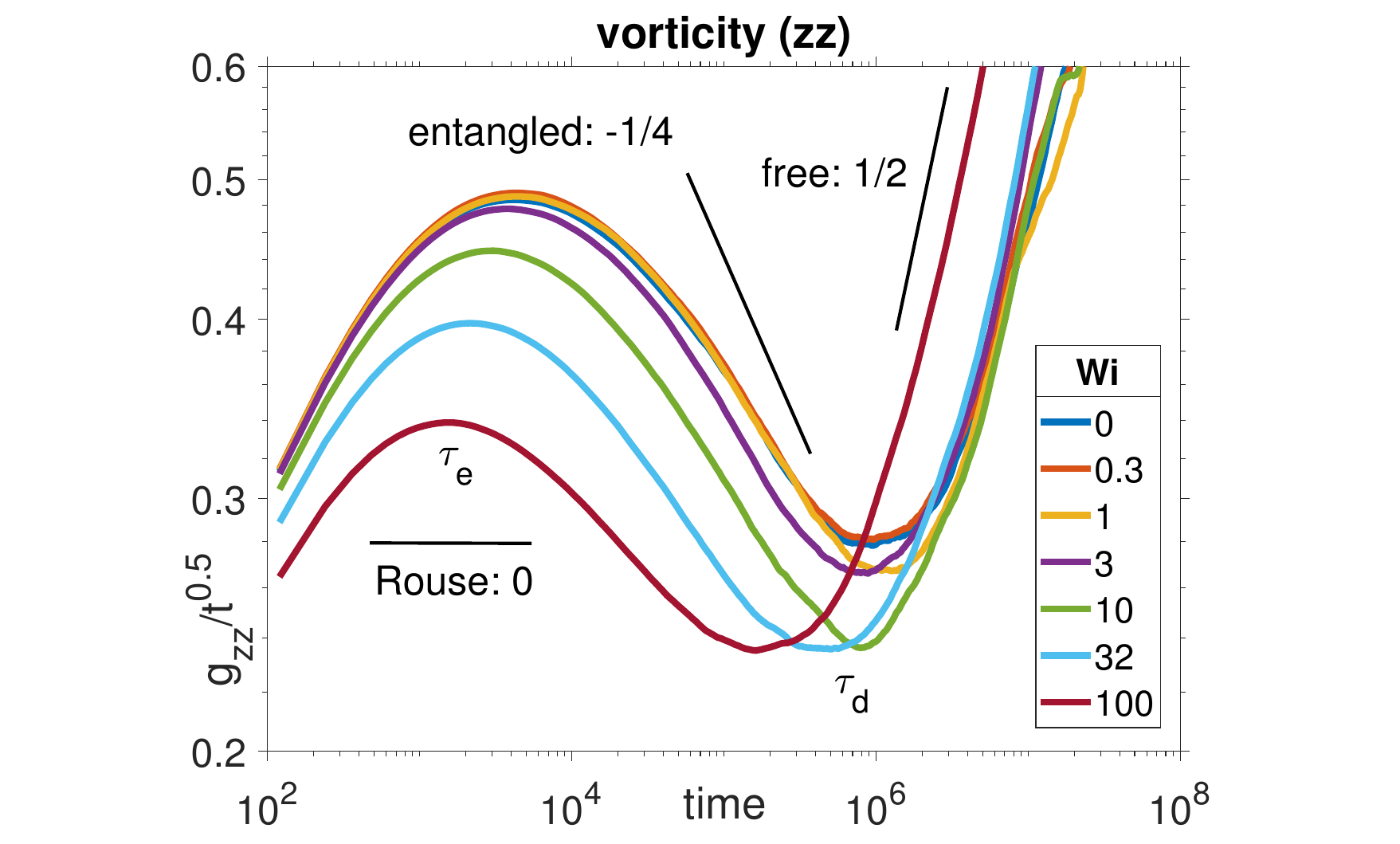}} 
			\includegraphics[clip,trim={.1\wd0} 0 {.13\wd0} 0,width=\linewidth]{./figures/gzz.pdf} 
		\endgroup
\end{subfigure}
\hfill
\begin{subfigure}{.49\textwidth}
\begingroup
			\sbox0{\includegraphics{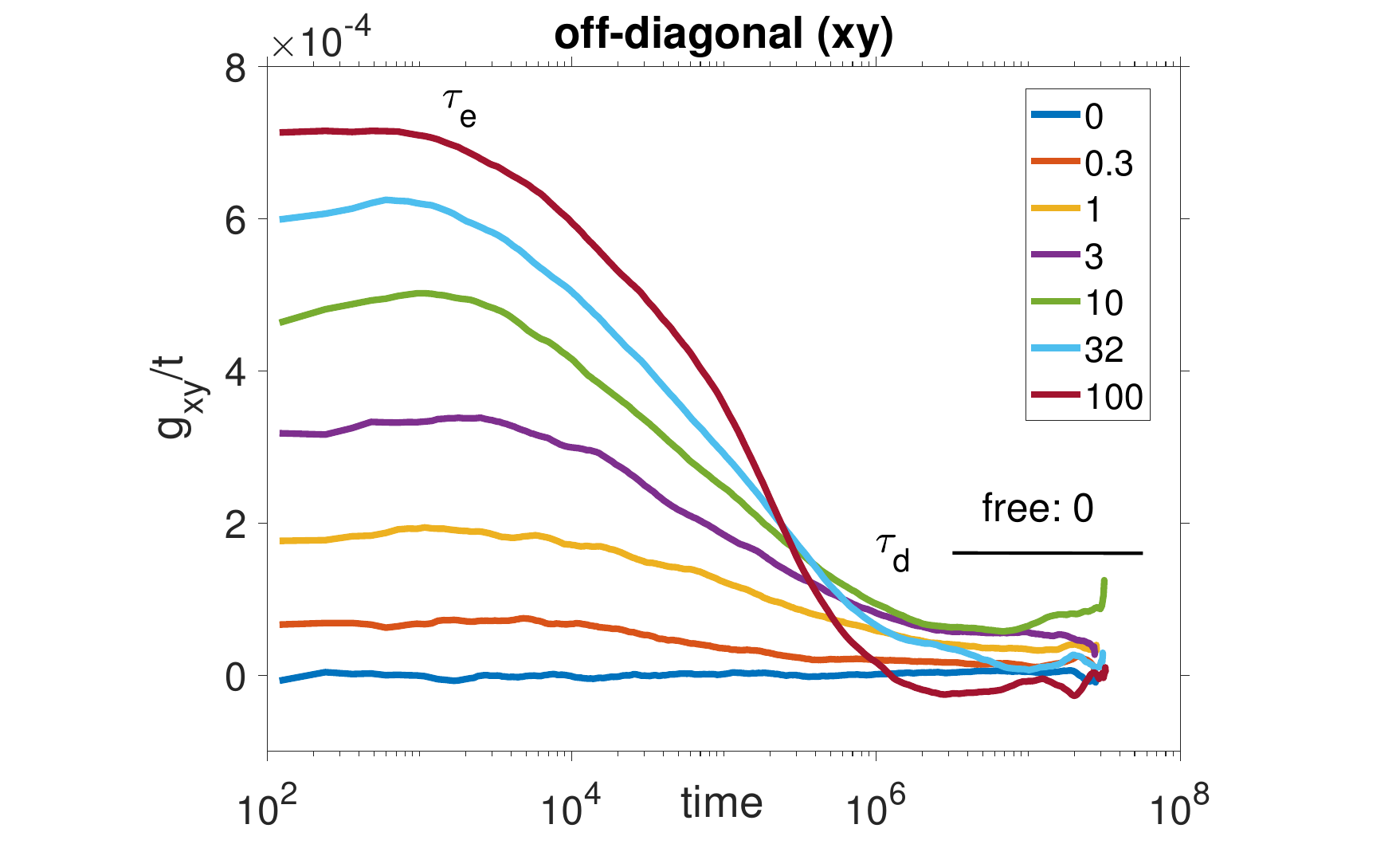}} 
			\includegraphics[clip,trim={.1\wd0} 0 {.13\wd0} 0,width=\linewidth]{./figures/gxy.pdf} 
		\endgroup
\end{subfigure}
\caption{Central monomer MSD at various shear rates Wi. Straight black lines show the equilibrium theoretical predictions according to reptation, Rouse, and free random walk models. The curves are divided by $t^{1/2}$ to reveal the entanglement $\tau_e$ and the disentanglement $\tau_d$ transition points. Under shear, the diffusion is seen to be altered differently along all three directions, in addition to being coupled in the $xy$ plane.}\label{g1}
\end{figure}

At high shear the system response becomes non-linear, and its topological network is disrupted as evidenced by a decreasing $Z$ number. This disentanglement trend was examined with primitive path analysis in Ref.~\cite{baig2010flow}. In the present study we aim to characterize the non-linear steady state from a dynamical standpoint. That is, we look at fluctuations around the deformed structure. Among the many quantities that could be considered, we choose diffusion, because it is clearly visible from the simulation perspective, and it can be measured experimentally with neutron spin echo (NSE)~\cite{kawecki2016probing}. The most topologically confined part of the molecule is the central monomer $\mathbf{R}(t,s=0.5) \equiv \mathbf{R}(t)$, and its diffusion under shear shows the strongest deviation from equilibrium. It is quantified by the mean square displacement (MSD), defined as
\begin{equation}
g_{\alpha \beta}(t) = \braket{[\mathbf{R}(t) - \mathbf{R}(0)]_{\alpha} [\mathbf{R}(t) - \mathbf{R}(0)]_{\beta}}
\end{equation}
plotted in Fig.~\ref{g1}. In the shear geometry there are four distinct components: $\alpha \beta = (xx,\,  yy,\, zz,\, xy)$, respectively referring to shear, gradient, vorticity, and the off-diagonal term in the shear-gradient plane. Concerning the diffusion along the $x$-axis, the raw trajectory is overwhelmed by a huge ballistic component $g_{xx}\propto t^2$ of the macroscopic shear flow, so we subtract it before the data analysis:
\begin{equation}\label{advection}
\mathbf{R}(t) \rightarrow \mathbf{R}(t) - \mathbf{e}_x \sum_{t'/\Delta t = 0}^{t/\Delta t} \left(\dot{\gamma}\Delta t\right) \mathbf{e}_y \cdot \mathbf{R}_{\text{cm}}(t')
\end{equation}
The big sum is the total displacement imparted on the center of mass (cm) of every chain, over the course of time from $0$ to $t$. In other words, we transform the trajectory to the oblique coordinate system where the effect of advection is removed (see Eq.~\eqref{covbasis}), retaining only the molecular diffusion.

In equilibrium, the MSD of long entangled chains is known to scale as $t^{1/4}$ (reptation), whereas unentangled Rouse chains follow $t^{1/2}$, while a random walk has $t^{1}$. To highlight the entanglement character, it is customary to divide the MSD by the Rouse law $t^{1/2}$, so that the topologically confined motion has a negative slope, as shown in Fig.~\ref{g1}. The negative portion of the curve is demarcated by two peaks, corresponding to the entanglement and the disentanglement times:
\begin{equation}\label{taue}
\tau_e = \frac{a^4 \zeta}{k_B T b^2}, \quad \tau_d = \frac{\zeta N^3 b^4}{\pi^2 k_B T a^2}
\end{equation}
The above formulas are derived for highly entangled polymers at equilibrium. The idea is that the monomer starts to feel the confining topological network only after having diffused a square distance $g_{\alpha \beta}(\tau_e) = a^2$, where $a$ is the tube radius. It then slowly escapes the initial network, and after traveling a square distance equal to the chain size, $g_{\alpha \beta}(\tau_d) = Nb^2$, it is fully relaxed in a new conformation.

\begin{figure}[ptbh!] 
\begin{subfigure}{.49\textwidth}
\begingroup
			\sbox0{\includegraphics{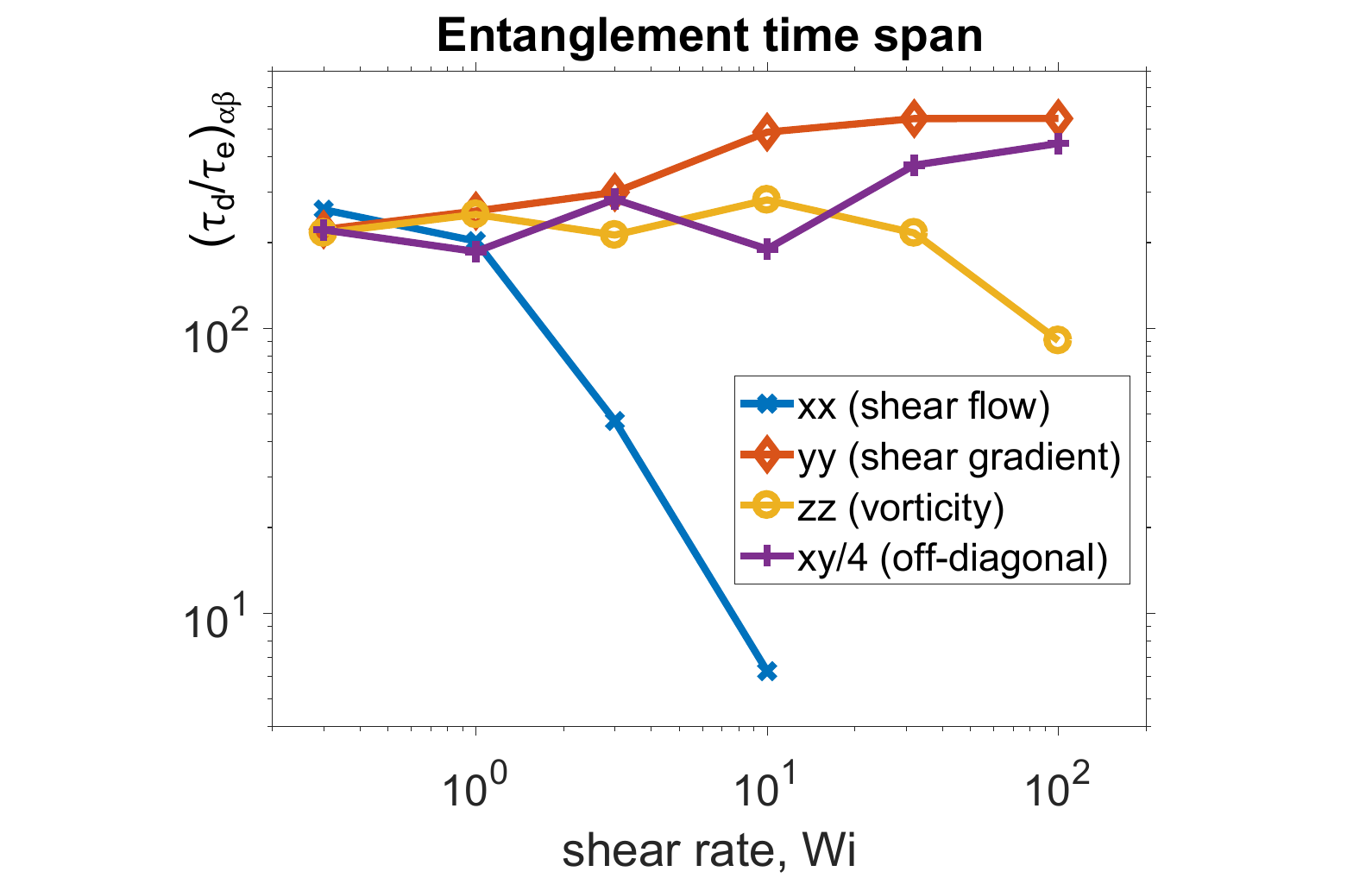}} 
			\includegraphics[clip,trim={.05\wd0} 0 {.15\wd0} 0,width=\linewidth]{./figures/etime.pdf} 
		\endgroup
\end{subfigure}
\hfill
\begin{subfigure}{.49\textwidth}
		\begingroup
			\sbox0{\includegraphics{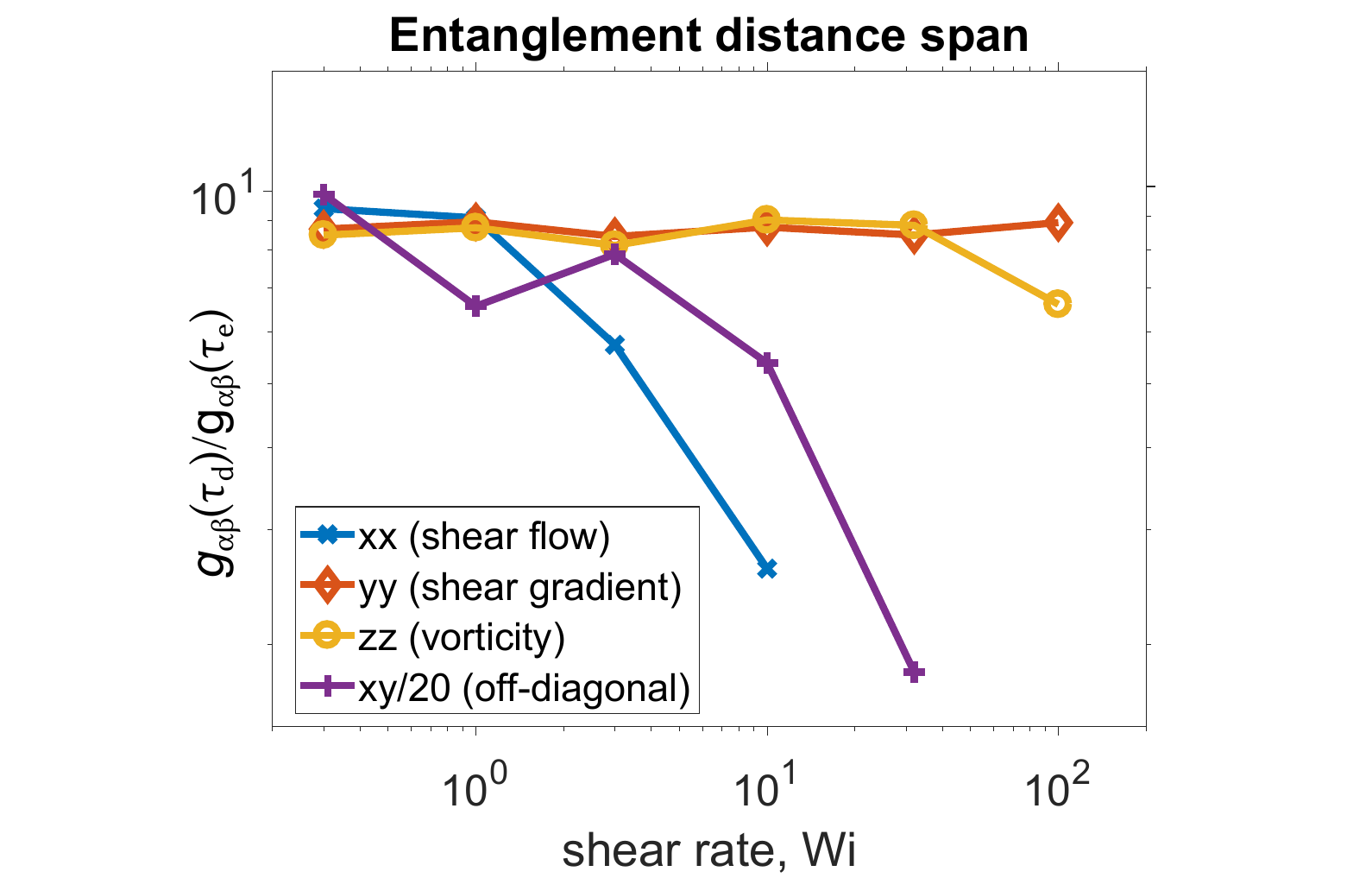}} 
			\includegraphics[clip,trim={.05\wd0} 0 {.15\wd0} 0,width=\linewidth]{./figures/edistance.pdf} 
		\endgroup
\end{subfigure}
\caption{A dynamical measure for the degree of entanglement. Left panel: the time required to escape topological confinement, right panel: the distance traveled during that time. Under shear, the polymer network is anisotropic, leading to different mobility in all three directions.}\label{edegree}
\end{figure}

While there is no consensus on how to extend Eqs.~\eqref{taue} to the non-equilibrium regime, it is clear from Fig.~\ref{g1} that the two scales $\tau_e$ and $\tau_d$ continue to exist and with increasing $\text{Wi}$ are gradually moving away from their equilibrium positions. We use these two dynamical peaks to quantify the extent of topological confinement in terms of time $(\tau_d/\tau_e)_{\alpha \beta}$ and MSD $g_{\alpha \beta}(\tau_d)/g_{\alpha \beta}(\tau_e)$ ratios, shown in Fig.~\ref{edegree}.

The largest effect of shear is along the flow direction $x$, where $g_{xx}(t)$ increases with shear at every timescale $t$, to the point where the entanglement character (the negative slope) disappears completely between $\text{Wi} = 3$ and $10$. An opposite, but weaker effect is found perpendicular to the flow. The shear gradient axis, $y$, shows a slowdown of dynamics at all timescales $t$. The ratio $(\tau_d/\tau_e)_{yy}$ (see Fig.~\ref{edegree}) reveals that the monomer remains topologically confined for up to five times longer along this axis. On the other hand, if we look at the ratio $g_{yy}(\tau_d)/g_{yy}(\tau_e)$, which quantifies the distance traveled to escape the confinement, it appears to be insensitive to shear. Next, let us inspect the cross-correlated displacement $g_{xy}(t)$ shown on the semilog plot in Fig.~\ref{g1}, normalized by $t$ to reveal its characteristic times $\tau_e$ and $\tau_d$. Curiously, the confinement time span $(\tau_d/\tau_e)_{xy}$ increases similarly to what is seen in the $g_{yy}$ component, while the distance span $g_{xy}(\tau_d)/g_{xy}(\tau_e)$ decreases rapidly just like in the $g_{xx}$ component.

The vorticity $z$ is perpendicular to the shear plane and shows the weakest effect. Here the change of dynamics is not monotonic, slowing down on a short time scale, but then speeding up to an extent where the terminal diffusion coefficient becomes slightly higher than in equilibrium. Noteworthy is that both the time $(\tau_d/\tau_e)_{zz}$ and the distance $g_{zz}(\tau_d)/g_{zz}(\tau_e)$ ratios remain roughly constant.

\begin{figure}[ptbh!] 
\begin{subfigure}{.49\textwidth}
		\begingroup
			\sbox0{\includegraphics{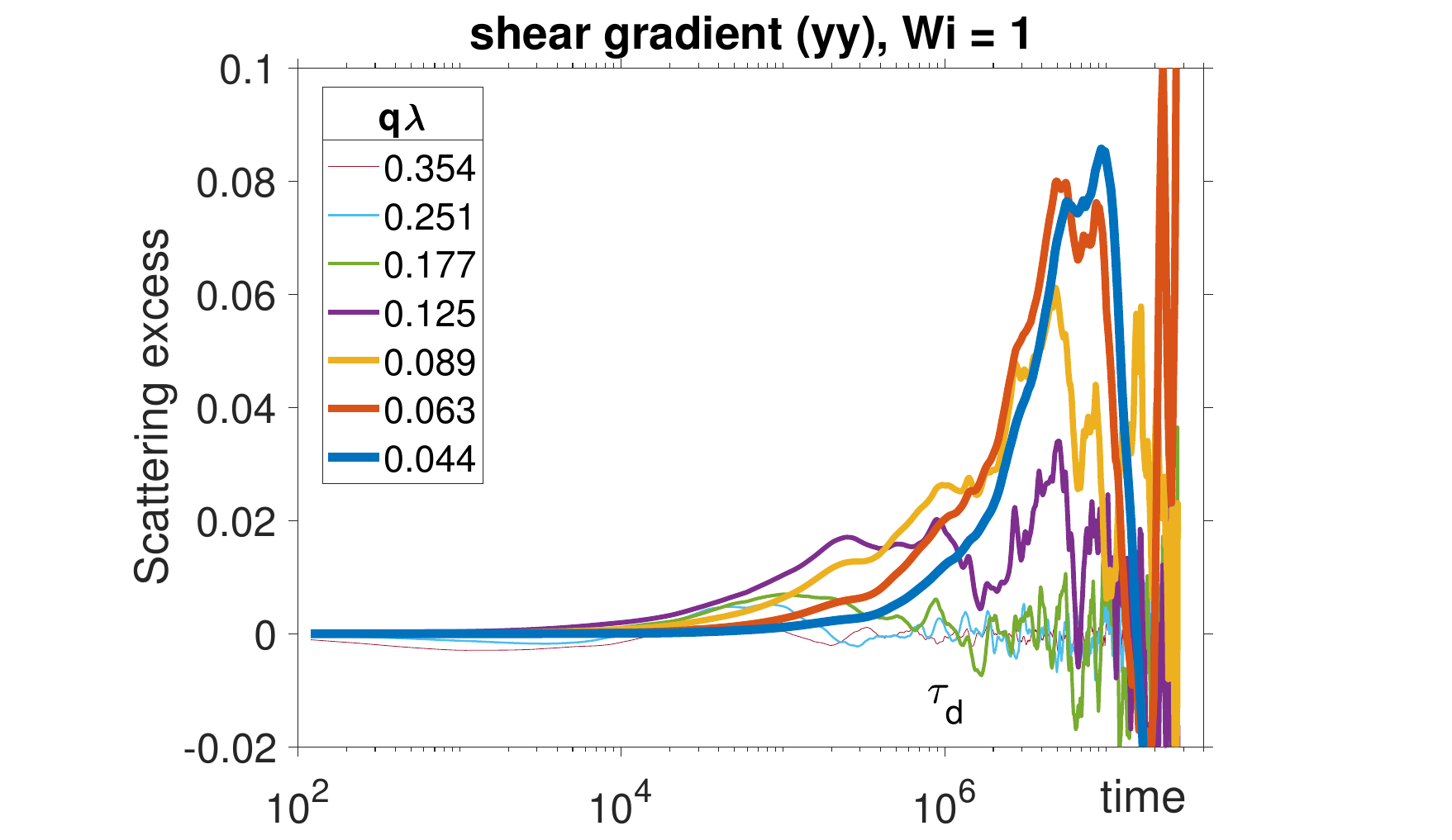}} 
			\includegraphics[clip,trim={.08\wd0} 0 {.14\wd0} 0,width=\linewidth]{./figures/NSEyy1.pdf} 
		\endgroup
\end{subfigure}
\hfill
\begin{subfigure}{.49\textwidth}
\begingroup
			\sbox0{\includegraphics{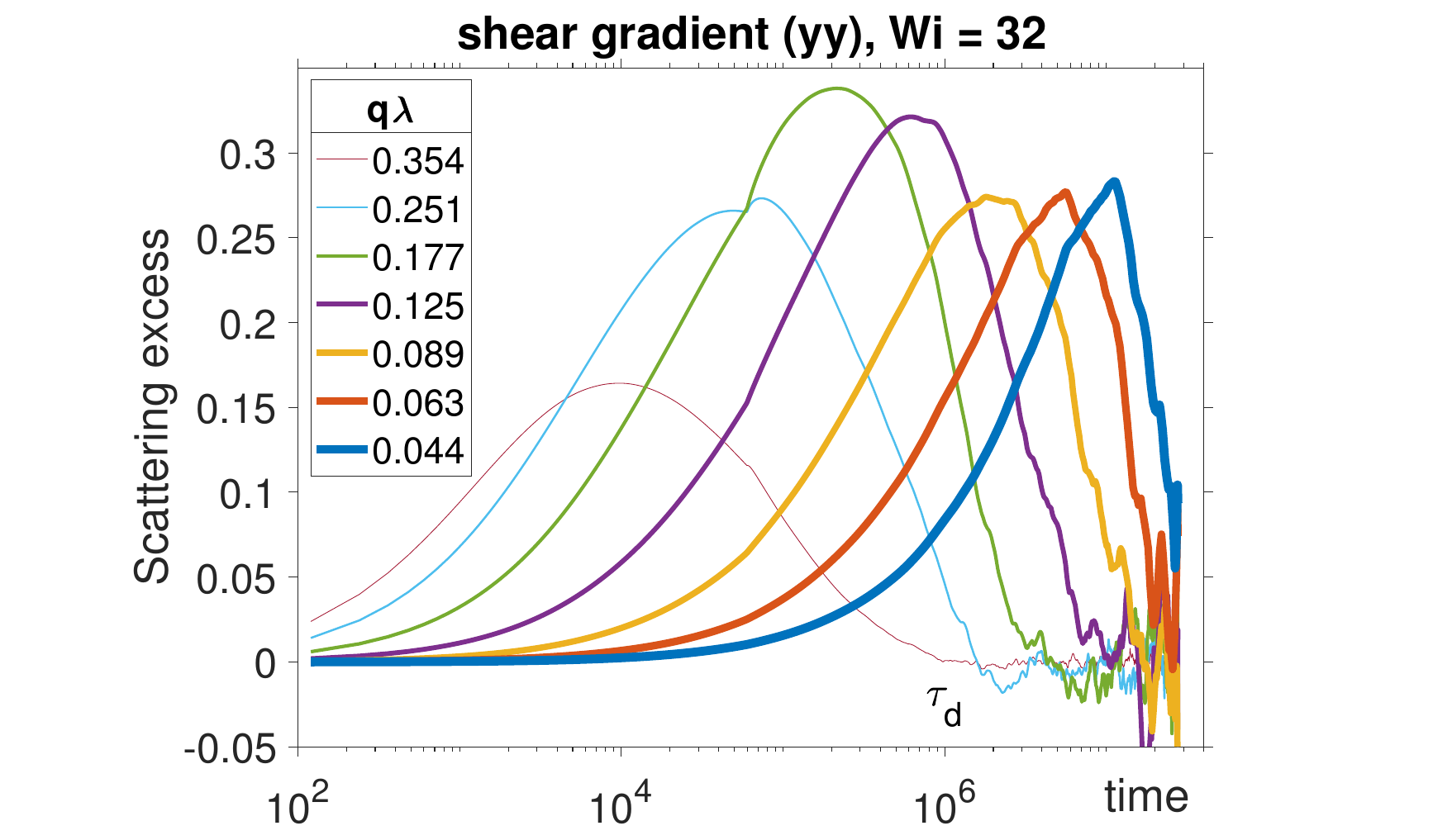}} 
			\includegraphics[clip,trim={.08\wd0} 0 {.14\wd0} 0,width=\linewidth]{./figures/NSEyy32.pdf} 
		\endgroup
\end{subfigure}
\vskip\baselineskip
\begin{subfigure}{.49\textwidth}
\begingroup
			\sbox0{\includegraphics{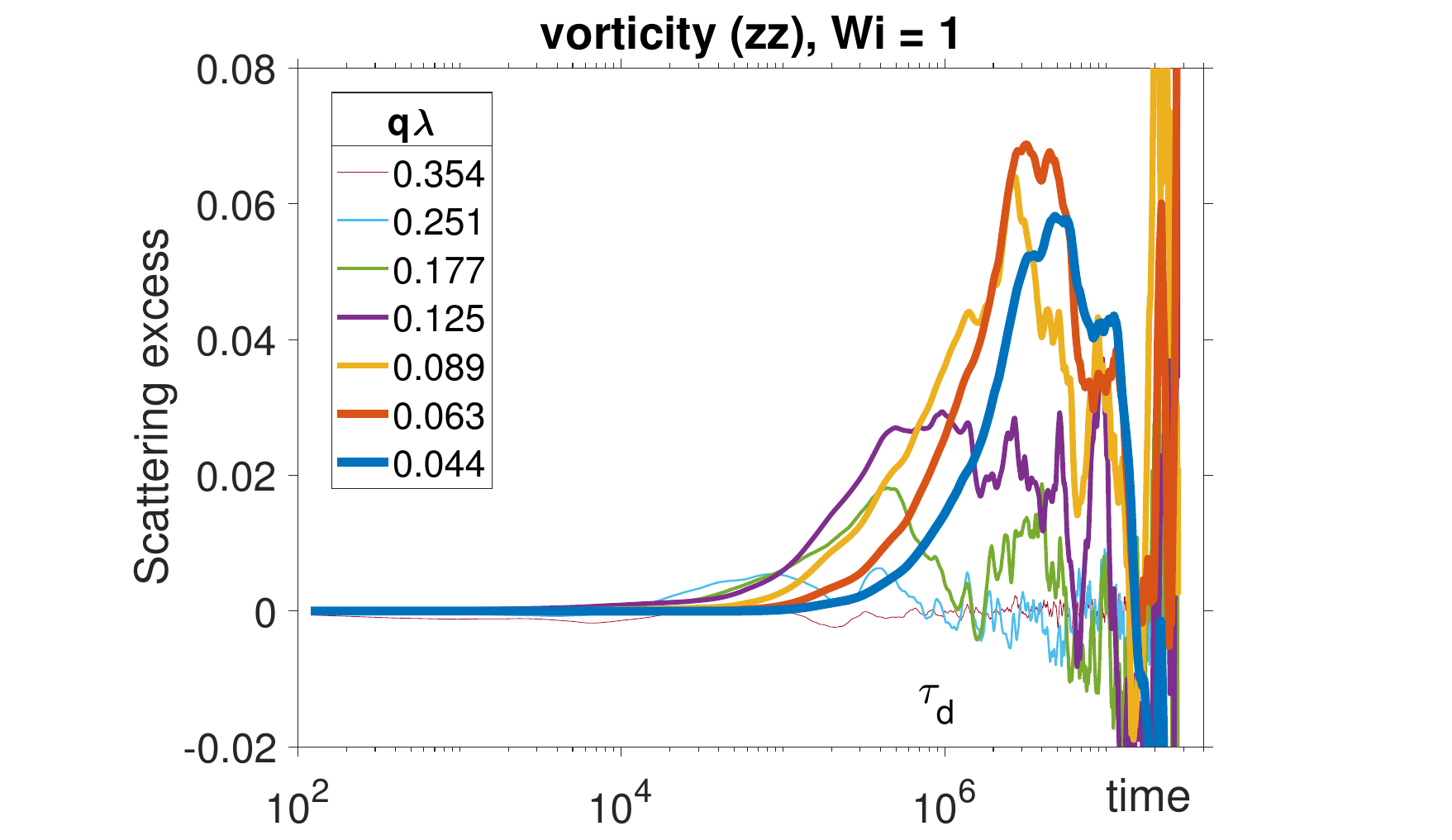}} 
			\includegraphics[clip,trim={.08\wd0} 0 {.14\wd0} 0,width=\linewidth]{./figures/NSEzz1.pdf} 
		\endgroup
\end{subfigure}
\hfill
\begin{subfigure}{.49\textwidth}
\begingroup
			\sbox0{\includegraphics{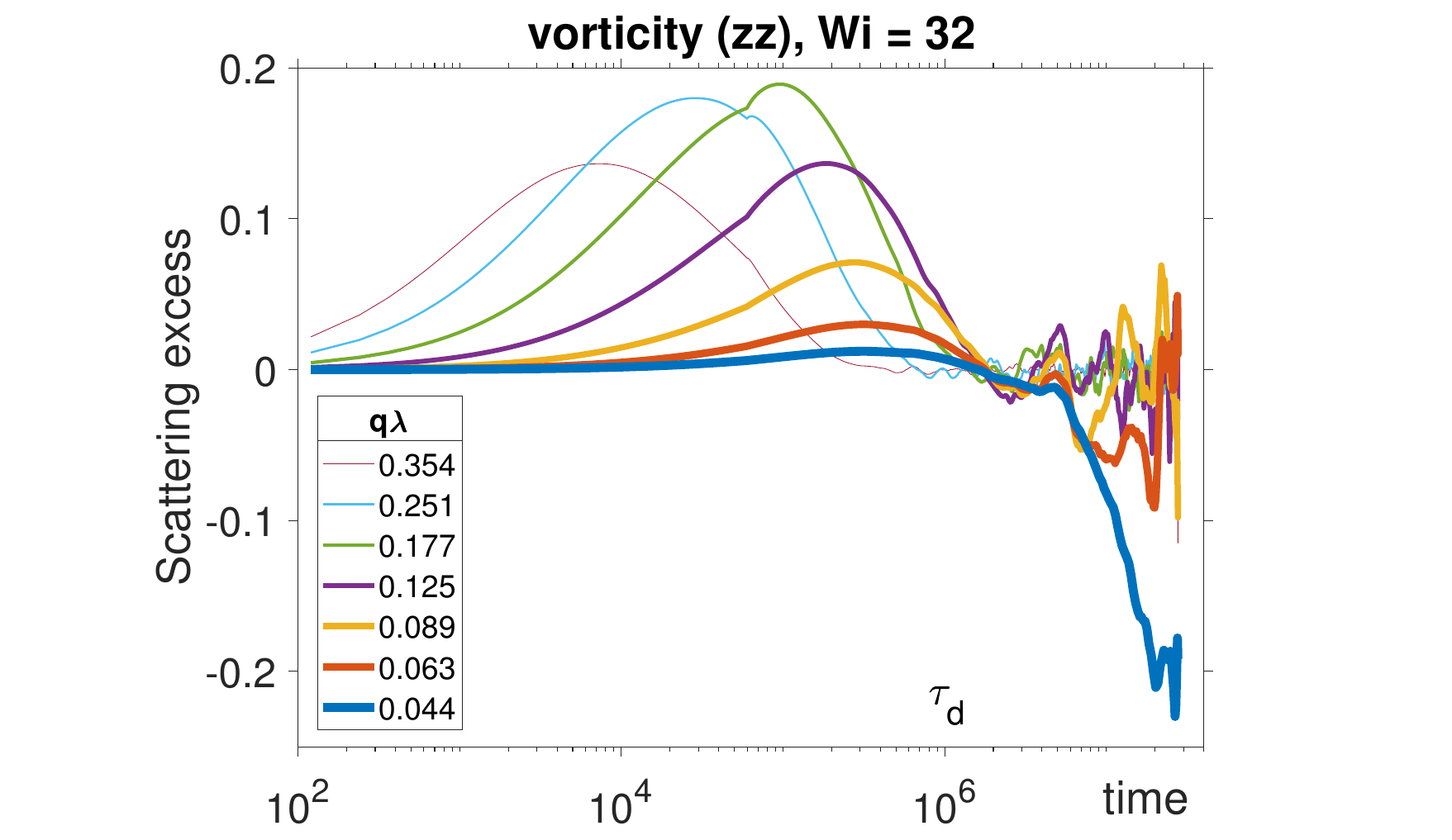}} 
			\includegraphics[clip,trim={.08\wd0} 0 {.14\wd0} 0,width=\linewidth]{./figures/NSEzz32.pdf} 
		\endgroup
\end{subfigure}
\caption{The intermediate scattering function excess under slow (left) and fast (right) shear, along the two directions which are accessible with rheo-NSE~\cite{kawecki2016probing}. At $\text{Wi}=1$ both axes show a slowdown (more entanglement), most pronounced at low $qR<1$ and long $t>\tau_d$. With increasing shear $\text{Wi}=32$, the slowdown shifts to higher $q$ and shorter $t$. While the gradient axis only has a slowdown, the vorticity axis eventually flips character and develops a speedup (less entanglement) at low $qR<1$ and long $t>\tau_d$.}\label{excess}
\end{figure}

Another way to visualize chain dynamics is the coherent intermediate scattering function:
\begin{equation}
F(\mathbf{q},t) = \frac{1}{J}\braket{\sum_{j,j'=1}^J \exp[ i\mathbf{q}\cdot (\mathbf{R}_{j}(t)-\mathbf{R}_{j'}(0))]}
\end{equation}
It can be measured with NSE and quantifies how likely are two monomers initially at a distance $1/q$ apart to stay correlated after a time $t$. A recent simulation~\cite{hsu2017detailed} has investigated this function in detail for strongly entangled polymers in equilibrium. Here we focus on the change that may be observed under shear flow, quantified by the scattering excess
\begin{equation}
E (t) = F_{\text{Wi}}(\mathbf{q},t)/F_{\text{Wi}}(\mathbf{q},0) - F_0(\mathbf{q},t)/F_0(\mathbf{q},0)
\end{equation}
which is the difference between the intermediate scattering function at rest and under shear, shown in Fig.~\ref{excess}. The excess $E(t)$ can range between $(-1, 1)$, where positive (negative) values indicate that there is a slowdown (speedup) of the dynamics. To reduce random noise, the raw data has been smoothed with a moving average filter of width $120000\Delta t = 0.12\tau_d$. We focus on the two axes perpendicular to shear, the $q_y$ and $q_z$, as they can be accessed by experiments like NSE and dielectric spectroscopy. A significant dynamical change begins at around $\text{Wi} = 1$, where both $y$ and $z$ components display a slowdown starting at low $qR \lesssim 1$ and long $t\gtrsim \tau_d$, the magnitude of the effect being \SI{50}{\percent} greater for the $y$-axis. (The small speedup visible at short $t$ is probably a simulation artifact due to numerical anisotropy imposed by the oblique grid, as explained in Appendix A). At a strong shear of $\text{Wi}=32$, the slowdown not only grows in magnitude, but also shifts to shorter time $t$, even four orders of magnitude shorter than the shear rate itself, demonstrating that a macroscopic shear flow can alter microscopic motion at the scale of barely one entanglement. While the $y$ component is slowed down at all times and distances, the situation is different for the $z$-axis. Its slowdown trend is reversed at around the disentanglement time $\tau_d$, where the low $q_z$-vectors start speeding up with respect to equilibrium.

\section{Discussion}
The largest structural change occurs in the flow direction $x$. To reveal the dynamics of those new structures, one must subtract the macroscopic shear flow, which is easier in simulation than experiment. Currently the only technique able to measure the hidden dynamics is the parallel superposition rheology. Although its data analysis is far from perfect~\cite{Unidad2014}, experiments on entangled PI solutions~\cite{Cleyn1981,vermant1998orthogonal} and more recently wormlike micelles~\cite{kim2013superposition}, both report a strong speedup of stress relaxation modulus, in qualitative agreement with our simulation results. 

The amplitude of structural change is smaller along the shear gradient $y$ and even smaller along the vorticity $z$. Nevertheless, those changes may be easier to detect experimentally, since they are not overwhelmed by the bulk flow. Dielectric spectroscopy can access the autocorrelation of the end-to-end vector $\Phi(t) = \braket{\mathbf{R}_{\text{ete}}(t) \cdot \mathbf{R}_{\text{ete}}(0)}$ of the polymer chain. Data on entangled PI solutions in the $y$-axis shows a slight speedup of relaxation dynamics, although the change is much smaller than predicted by tube theory~\cite{watanabe2003rheo}. Even though we have not analyzed the simulated end-to-end vector separately, the experimental trend of overall speedup disagrees with our simulation, since we found a clear slowdown for all $q_y$ vectors.

Another tool to probe polymer dynamics under shear is NMR spectroscopy. It can access intramolecular dynamics, corresponding to an entanglement strand, or high $q$ region. A slowdown of dynamics in the shear gradient direction $y$ was reported~\cite{bohme2011nmr}, this time in accord with our simulation data.

Recently, neutron spin echo experiments under \emph{in situ} shear became possible~\cite{kawecki2016probing}, measuring the intermediate scattering function $F(q_z,t)$ along the $z$ (vorticity) axis. A PDMS melt sheared at $\text{Wi} = 0.4$ showed no change in scattering~\cite{Kawecki2017}. Although higher shear rates could not be reached at this stage, the null result agrees with our simulation, which shows only minuscule change below $\text{Wi} = 1$.

Labelled polymers can be filmed under a microscope to visualize their motion in real space. Observations of flowing entangled DNA~\cite{teixeira2007individualistic} and semiflexible actin filaments~\cite{harasim2013direct} have been described in terms of tumbling dynamics, which are abrupt fluctuations between strongly elongated and collapsed states. This can be understood with a double potential well, where the chain spends most of the time confined by the flowing network, but can occasionally flip around by \SI{180}{\degree}, where it finds another elongated state of an equal probability. The jump must be rapid since being misaligned with the flow is energetically unfavorable. We believe that this tumbling behaviour is reflected in our simulated scattering excess along the $z$ axis (Fig.~\ref{excess}). The elongated chain state corresponds to the slowdown seen at short $t$ and high $q$ (local confinement), while the occasional rapid jumps are seen as faster dynamics at long $t$ and low $q$ (global speedup). In agreement with our simulation, a clear speedup along the $z$ axis is also found in superposition rheology applied perpendicular to shear flow~\cite{Cleyn1981,vermant1998orthogonal,kim2013superposition}, which is sensitive to molecular motion at long time and distance scales.

To conclude, the simulation results are in qualitative agreement with most of the dynamics experiments conducted under shear so far.  We are looking forward to the emerging technique of rheo-NSE spectroscopy to verify the local slowdown for the $z$-axis as predicted by the simulation.

\section{Acknowledgements}
The authors thank Jean-Louis Barrat for his valuable advice. Gunnar K. Palsson, Anton Devishvili, and Franz A. Adlmann have contributed to the development of the simulation software. The Carl-Tryggers stiftelse as well as the Swedish research council are acknowledged for financial support, grant CTS 16:519. The Quadro P5000 used for this research was donated by the NVIDIA Corporation. 

\section{Appendix A. Oblique coordinate system}
\begin{figure}[bht]
\centering
		\begingroup
			\sbox0{\includegraphics{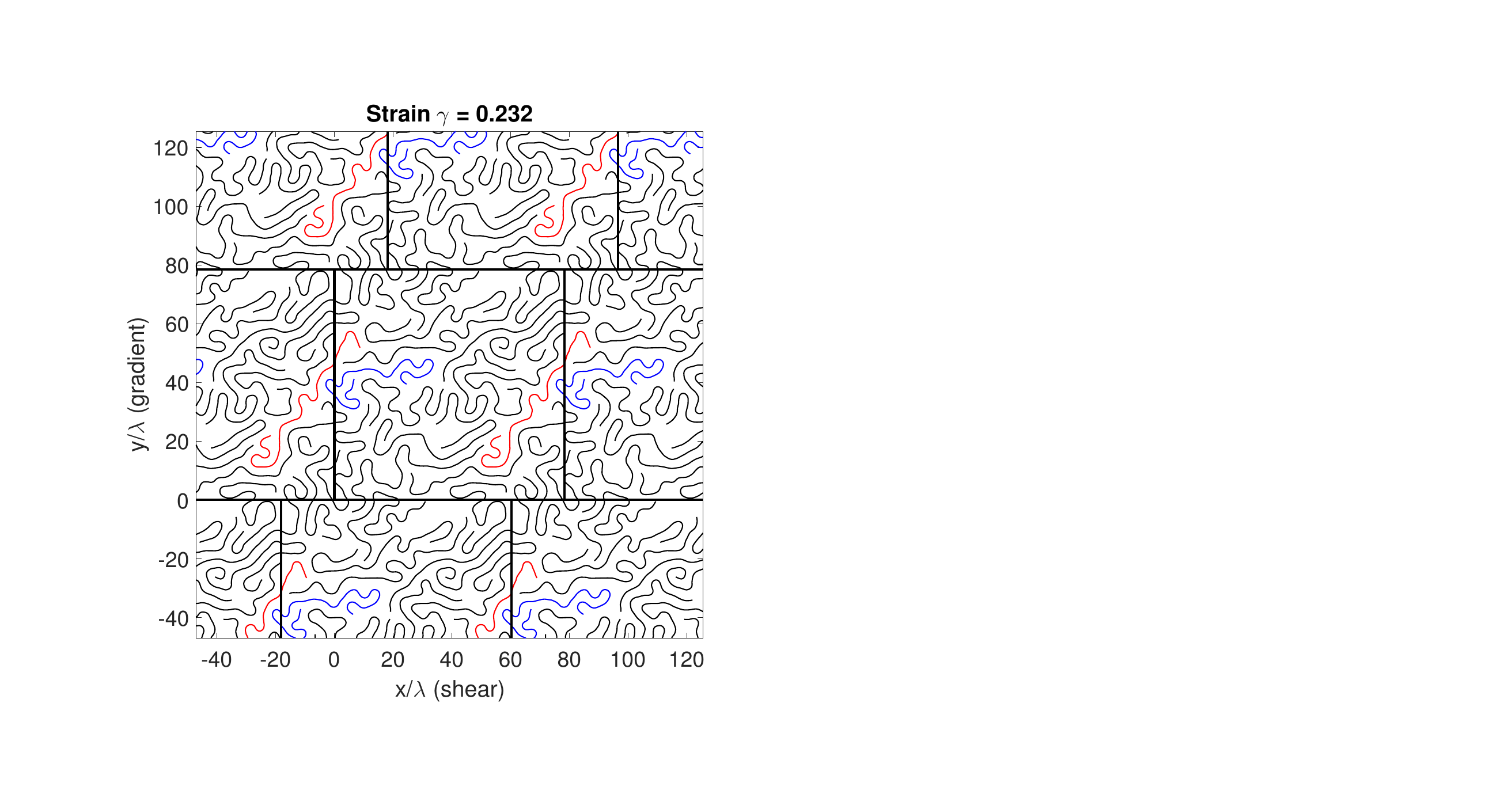}} 
			\includegraphics[clip,trim={.0\wd0} 0 {.5\wd0} 0,width=0.7\linewidth]{./figures/oblique.pdf} 
		\endgroup
    \caption{A two-dimensional example of shear, using Lees-Edwards boundary conditions. Two random chains are colored for clarity.}\label{leesedwards}
\end{figure}
\begin{figure}[bht]
\centering
		\begingroup
			\sbox0{\includegraphics{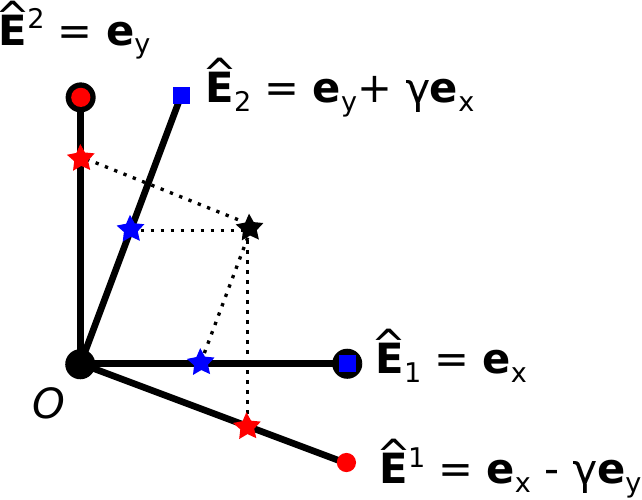}} 
			\includegraphics[clip,trim={.0\wd0} 0 {.0\wd0} 0,width=0.5\linewidth]{./figures/oblique_system.pdf} 
		\endgroup
    \caption{The definition of an oblique coordinate system of some arbitrary strain $\gamma=0.375$ in terms of the rectangular basis $(\mathbf{e}_x, \mathbf{e}_y)$. The oblique system has two bases, the covariant $(\mathbf{\hat{E}}_1, \mathbf{\hat{E}}_2)$ and the contravariant $(\mathbf{\hat{E}}^1, \mathbf{\hat{E}}^2)$. A particle position $\mathbf{r}$ shown by a black star is decomposed into its covariant and contravariant components.}\label{obliquesystem}
\end{figure}

To simulate bulk shear flow, we need to impose special boundary conditions, invented by Lees and Edwards, as illustrated in Fig.~\ref{leesedwards}. A further complication is that we need to use the Fourier transform to speed up the excluded volume force calculation, and may also need it to evaluate the polymer structure factor for comparison with SANS. The technique was first developed in Ref.~\cite{kobayashi2011implementation}, and here we adapt their approach to our situation.

In a cubic box of length $L$, every particle $\mathbf{R}$ is replicated at points
\begin{equation}
\mathbf{R} \rightarrow \mathbf{R} + L[(\gamma n_y + n_x)\mathbf{e}_x + n_y \mathbf{e}_y + n_z \mathbf{e}_z],
\end{equation}
where the $n_x$, $n_y$, and $n_z$ are integers $-\infty,\, \ldots,\, -1,\, 0,\, +1\, \ldots,\, +\infty$. The box strain
\begin{equation}
\gamma(t) = \int_{-\infty}^t \dot{\gamma}(t')\,dt'
\end{equation}
is incremented at every time step, and the oblique-periodic images can shift many times over their own length during a long run. For numerical stability reasons, the strain is always folded to
\begin{equation}
\gamma \rightarrow \gamma - \texttt{round}(\gamma)
\end{equation}
restricting it to the symmetric range $-0.5<\gamma<0.5$.

A particle position in the rectangular coordinate system $(\mathbf{e}_x, \mathbf{e}_y, \mathbf{e}_z)$ can be decomposed into its Cartesian components $(x,y,z)$:
\begin{equation}
\mathbf{r} = x\mathbf{e}_x + y\mathbf{e}_y + z\mathbf{e}_z.
\end{equation}
In terms of the rectangular unit vectors, the oblique coordinate system of strain $\gamma$ is defined by the covariant basis:
\begin{subequations}
\begin{align}
\hat{\mathbf{E}}_1 &= \mathbf{e}_x\\
\hat{\mathbf{E}}_2 &= \gamma \mathbf{e}_x + \mathbf{e}_y\\
\hat{\mathbf{E}}_3 &= \mathbf{e}_z
\end{align}
\end{subequations}
In other words, the metric tensor of the oblique system is given by
\begin{equation}
g_{ij} = \hat{\mathbf{E}}_i \cdot \hat{\mathbf{E}}_j = \begin{pmatrix}
1	& \gamma	&	0\\
\gamma	& 1+\gamma^2	& 0\\
0 & 0 & 1
\end{pmatrix}
\end{equation}
As illustrated in Fig.~\ref{obliquesystem}, the position of a particle in the oblique system, denoted by $\hat{\mathbf{r}}$, is expressed as
\begin{equation}
\hat{\mathbf{r}} = \hat{x}^1 \hat{\mathbf{E}}_1 + \hat{x}^2 \hat{\mathbf{E}}_2 + \hat{x}^3 \hat{\mathbf{E}}_3
\end{equation}
where its coordinate components in the oblique system can be found by solving the equation $\hat{\mathbf{r}} = \mathbf{r}$:
\begin{subequations}\label{covbasis}
\begin{align}
\hat{x}^1 &= x- \gamma y\\
\hat{x}^2 &= y\\
\hat{x}^3 &= z
\end{align}
\end{subequations}
The simulated polymer network is periodic in these coordinates. The inverse is
\begin{subequations}\label{inversetransformation}
\begin{align}
x &= \hat{x}^1 + \gamma \hat{x}^2\\
y &= \hat{x}^2\\
z &= \hat{x}^3
\end{align}
\end{subequations}

The interaction potential is a scalar quantity, in our case a Gaussian:
\begin{equation}
U(\mathbf{r}) = \exp\left[-\frac{\mathbf{r}^2}{2\lambda^2}\right] = \exp \left[-\frac{\left(x^2+y^2+z^2\right)}{2\lambda^2}\right].
\end{equation}
A scalar remains invariant regardless of a coordinate system, so to express it in the oblique coordinates, we simply plug in the values from Eq.~\eqref{inversetransformation}:
\begin{equation}
\hat{U}(\hat{\mathbf{r}}) = \exp \left[-\frac{\left(\hat{x}^1+\gamma \hat{x}^2\right)^2 + \left(\hat{x}^2\right)^2 + \left(\hat{x}^3\right)^2}{2\lambda^2}\right]
\end{equation}
The force, in the oblique system, is given by the gradient:
\begin{equation}
\hat{\mathbf{f}}(\hat{\mathbf{r}}) = -\hat{\nabla} \hat{U}(\mathbf{r}) = -\left[\hat{\mathbf{E}}^1 \frac{\partial \hat{U}}{\partial \hat{x}^1} + \hat{\mathbf{E}}^2 \frac{\partial \hat{U}}{\partial \hat{x}^2} + \hat{\mathbf{E}}^3 \frac{\partial \hat{U}}{\partial \hat{x}^3} \right]
\end{equation}
which is expanded in terms of the contravariant basis vectors:
\begin{subequations}\label{contravariant}
\begin{align}
\hat{\mathbf{E}}^1 &= \mathbf{e}_x - \gamma \mathbf{e}_y\\
\hat{\mathbf{E}}^2 &= \mathbf{e}_y\\
\hat{\mathbf{E}}^3 &= \mathbf{e}_z
\end{align}
\end{subequations}
These are defined to be conjugate to the covariant basis vectors: $\hat{\mathbf{E}}_i \cdot \hat{\mathbf{E}}^j = \delta_i^j$, and in particular $\hat{\mathbf{E}}_2 \cdot \hat{\mathbf{E}}^1 = (\gamma \mathbf{e}_x + \mathbf{e}_y) \cdot (\mathbf{e}_x - \gamma \mathbf{e}_y) = 0$. Morevover, the contravariant metric tensor
\begin{equation}
g^{ij} = \hat{\mathbf{E}}^i \cdot \hat{\mathbf{E}}^j = \begin{pmatrix}
1+\gamma^2	& -\gamma	&	0\\
-\gamma	& 1	& 0\\
0 & 0 & 1
\end{pmatrix}
\end{equation}
is clearly the inverse of the covariant one: $g_{ij}g^{ij} = 1$. Let us be very clear on what are the specific components of the force, in terms of the oblique contravariant basis:
\begin{subequations}
\begin{align}
\hat{f}_1 &= -\frac{\partial \hat{U}}{\partial \hat{x}^1} = \left(\frac{\hat{x}^1 + \gamma \hat{x}^2}{\lambda^2}\right) \hat{U}(\hat{\mathbf{r}})\\
\hat{f}_2 &= -\frac{\partial \hat{U}}{\partial \hat{x}^2} = \left(\frac{\gamma \hat{x}^1 + \left(1+\gamma^2\right)\hat{x}^2}{\lambda^2}\right) \hat{U}(\hat{\mathbf{r}})\\ \label{extraterms}
\hat{f}_3 &= -\frac{\partial \hat{U}}{\partial \hat{x}^3} = \left(\frac{\hat{x}^3}{\lambda^2}\right) \hat{U}(\hat{\mathbf{r}})
\end{align}
\end{subequations}

Just like the position vector, the force vector can be expanded in either the rectangular or the oblique systems:
\begin{subequations}
\begin{align}
\mathbf{f} &= f_x \mathbf{e}_x + f_y \mathbf{e}_y + f_z \mathbf{e}_z\\
 &= \hat{f}_1 \hat{\mathbf{E}}^1 + \hat{f}_2 \hat{\mathbf{E}}^2 + \hat{f}_3 \hat{\mathbf{E}}^3
\end{align}
\end{subequations}
which leads to its rectangular components
\begin{subequations}
\begin{align}
f_x &= \hat{f}_1\\
f_y &= \hat{f}_2 - \gamma \hat{f}_1\\
f_z &= \hat{f}_3
\end{align}
\end{subequations}

Having presented all the necessary formulas, we now outline the algorithm required to evaluate the excluded volume force in the oblique periodic system.
\begin{enumerate}
\item We start off with a set of all particle positions $(x,y,z)$ given in the usual rectangular coordinates. A fixed strain value $\gamma$ is given as well.
\item The positions are transformed into the oblique system: $(x,y,z)\rightarrow (x-\gamma y, y, z)$, and the periodic boundary conditions are applied: $\hat{\mathbf{r}} = \hat{\mathbf{r}} - L\, \texttt{floor}\, (\hat{\mathbf{r}}/L)$. We are now folded inside the central oblique box, which is periodic along its covariant unit vectors.
\item A density histogram $\hat{\rho}(\hat{\mathbf{r}})$ is constructed, which holds the number of particles in every oblique pixel $\hat{\mathbf{r}} = L\left(k_1 \hat{\mathbf{E}}^1 + k_2 \hat{\mathbf{E}}^2 + k_3 \hat{\mathbf{E}}^3\right)$ where the indices run from $k = 1,\, 2,\, \ldots, K = \texttt{round}(L/\Delta x)$, and $\Delta x$ is the pixel width.
\item The density is Fourier-transformed into the oblique reciprocal space:
\begin{equation}
\hat{\rho}(\hat{\mathbf{q}}) = \sum_{\hat{\mathbf{r}}} \hat{\rho}(\hat{\mathbf{r}}) e^{-i\mathbf{\hat{r}}\cdot \mathbf{\hat{q}}}
\end{equation}
Notice that the components of the reciprocal lattice vector $\hat{\mathbf{q}} = (\hat{q}_1, \hat{q}_2, \hat{q}_3)$ come out in the contravariant basis defined in Eq.~\eqref{contravariant}, because this basis has the unique property that
\begin{equation}
\mathbf{\hat{r}}\cdot \mathbf{\hat{q}} = \left(\hat{x}^1 \mathbf{\hat{E}}_1 + \hat{x}^2 \mathbf{\hat{E}}_2 + \hat{x}^3 \mathbf{\hat{E}}_3\right) \cdot \left(\hat{q}_1 \mathbf{\hat{E}}^1 + \hat{q}_2 \mathbf{\hat{E}}^2 + \hat{q}_3 \mathbf{\hat{E}}^3\right) = \hat{x}^1 \hat{q}_1 + \hat{x}^2 \hat{q}_2 + \hat{x}^3 \hat{q}_3
\end{equation}
\item Steps 3-4 are repeated for the force kernel. First, the force array $\mathbf{\hat{f}}(\mathbf{\hat{r}})$ is declared in the real oblique space, and then it is Fourier-transformed to the reciprocal oblique space to give $\mathbf{\hat{f}}(\mathbf{\hat{q}})$.
\item The force on every oblique pixel $\mathbf{\hat{r}}$ is found by the convolution theorem:
\begin{equation}
\mathbf{\hat{F}}(\mathbf{\hat{r}}) = \texttt{IFFT} [\hat{\rho}(\mathbf{\hat{q}}) \cdot \mathbf{\hat{f}}(\mathbf{\hat{q}})]
\end{equation}
\item Lastly, the oblique force is transformed to the rectangular system (the inverse of step 2, but for the contravariant basis): $(\hat{F}_1, \hat{F}_2, \hat{F}_3) \rightarrow (\hat{F}_1, \hat{F}_2 - \gamma \hat{F}_1, \hat{F}_3) = (F_x,F_y,F_z)$. These force components are then used in the equation of motion, Eq.~\eqref{eqmotion}, solved in the rectangular system as usual.
\end{enumerate}

The strain position is updated at every time step, $\gamma \rightarrow \gamma + \dot{\gamma}\Delta t$, and wrapped periodically: $\gamma \rightarrow \gamma - \texttt{round}(\gamma)$. In addition, it must be noted that the use of the discrete convolution introduces some degree of numerical anisotropy due to the finite size of the pixels. In equilibrium, the pixels are rectangular, and the numerical anisotropy is symmetric around every axis, which does not result in a net bias. When the pixels are oblique, however, the anisotropy tilts and produces a net numerical stress, which is unphysical and is not related to the shear flow. In fact, the numerical stress would persist even at equilibrium, if we would choose to simulate a system in a fixed oblique periodic system of $\gamma \neq 0$. To mitigate this parasitic effect, we keep track of the total accumulated strain
\begin{equation}\label{mitigation}
\gamma_{\text{sum}} = \sum \gamma(t)
\end{equation}
and for every new timestep we choose the image of $\gamma$ which would bring the accumulated strain $\gamma_{\text{sum}}$ as close to zero as possible. For instance, in the example of Fig.~\ref{leesedwards}, the strain is $\gamma = 0.232$, but one could equally well describe the system using a strain $\gamma^* = \gamma-1 = -0.768$. Therefore, our code propagates two timesteps using the value $\gamma$, and the third step would use the conjugate $\gamma^*$, because $|3\gamma-1|<|3\gamma|$.



\bibliography{manuscript}

\begin{thebibliography}{73}%
\makeatletter
\providecommand \@ifxundefined [1]{%
 \@ifx{#1\undefined}
}%
\providecommand \@ifnum [1]{%
 \ifnum #1\expandafter \@firstoftwo
 \else \expandafter \@secondoftwo
 \fi
}%
\providecommand \@ifx [1]{%
 \ifx #1\expandafter \@firstoftwo
 \else \expandafter \@secondoftwo
 \fi
}%
\providecommand \natexlab [1]{#1}%
\providecommand \enquote  [1]{``#1''}%
\providecommand \bibnamefont  [1]{#1}%
\providecommand \bibfnamefont [1]{#1}%
\providecommand \citenamefont [1]{#1}%
\providecommand \href@noop [0]{\@secondoftwo}%
\providecommand \href [0]{\begingroup \@sanitize@url \@href}%
\providecommand \@href[1]{\@@startlink{#1}\@@href}%
\providecommand \@@href[1]{\endgroup#1\@@endlink}%
\providecommand \@sanitize@url [0]{\catcode `\\12\catcode `\$12\catcode
  `\&12\catcode `\#12\catcode `\^12\catcode `\_12\catcode `\%12\relax}%
\providecommand \@@startlink[1]{}%
\providecommand \@@endlink[0]{}%
\providecommand \url  [0]{\begingroup\@sanitize@url \@url }%
\providecommand \@url [1]{\endgroup\@href {#1}{\urlprefix }}%
\providecommand \urlprefix  [0]{URL }%
\providecommand \Eprint [0]{\href }%
\providecommand \doibase [0]{http://dx.doi.org/}%
\providecommand \selectlanguage [0]{\@gobble}%
\providecommand \bibinfo  [0]{\@secondoftwo}%
\providecommand \bibfield  [0]{\@secondoftwo}%
\providecommand \translation [1]{[#1]}%
\providecommand \BibitemOpen [0]{}%
\providecommand \bibitemStop [0]{}%
\providecommand \bibitemNoStop [0]{.\EOS\space}%
\providecommand \EOS [0]{\spacefactor3000\relax}%
\providecommand \BibitemShut  [1]{\csname bibitem#1\endcsname}%
\let\auto@bib@innerbib\@empty
\bibitem [{\citenamefont {Cho}, \citenamefont {Ahn},\ and\ \citenamefont
  {Lee}(2004)}]{cho2004universality}%
  \BibitemOpen
  \bibfield  {author} {\bibinfo {author} {\bibfnamefont {K.~S.}\ \bibnamefont
  {Cho}}, \bibinfo {author} {\bibfnamefont {K.~H.}\ \bibnamefont {Ahn}}, \ and\
  \bibinfo {author} {\bibfnamefont {S.~J.}\ \bibnamefont {Lee}},\ }\href@noop
  {} {\bibfield  {journal} {\bibinfo  {journal} {Journal of Polymer Science
  Part B: Polymer Physics}\ }\textbf {\bibinfo {volume} {42}},\ \bibinfo
  {pages} {2730} (\bibinfo {year} {2004})}\BibitemShut {NoStop}%
\bibitem [{\citenamefont {Cavaille}\ \emph {et~al.}(1987)\citenamefont
  {Cavaille}, \citenamefont {Jourdan}, \citenamefont {Perez}, \citenamefont
  {Monnerie},\ and\ \citenamefont {Johari}}]{cavaille1987time}%
  \BibitemOpen
  \bibfield  {author} {\bibinfo {author} {\bibfnamefont {J.}~\bibnamefont
  {Cavaille}}, \bibinfo {author} {\bibfnamefont {C.}~\bibnamefont {Jourdan}},
  \bibinfo {author} {\bibfnamefont {J.}~\bibnamefont {Perez}}, \bibinfo
  {author} {\bibfnamefont {L.}~\bibnamefont {Monnerie}}, \ and\ \bibinfo
  {author} {\bibfnamefont {G.}~\bibnamefont {Johari}},\ }\href@noop {}
  {\bibfield  {journal} {\bibinfo  {journal} {Journal of Polymer Science Part
  B: Polymer Physics}\ }\textbf {\bibinfo {volume} {25}},\ \bibinfo {pages}
  {1235} (\bibinfo {year} {1987})}\BibitemShut {NoStop}%
\bibitem [{\citenamefont {Kasaai}, \citenamefont {Charlet},\ and\ \citenamefont
  {Arul}(2000)}]{kasaai2000master}%
  \BibitemOpen
  \bibfield  {author} {\bibinfo {author} {\bibfnamefont {M.~R.}\ \bibnamefont
  {Kasaai}}, \bibinfo {author} {\bibfnamefont {G.}~\bibnamefont {Charlet}}, \
  and\ \bibinfo {author} {\bibfnamefont {J.}~\bibnamefont {Arul}},\ }\href@noop
  {} {\bibfield  {journal} {\bibinfo  {journal} {Food Research International}\
  }\textbf {\bibinfo {volume} {33}},\ \bibinfo {pages} {63} (\bibinfo {year}
  {2000})}\BibitemShut {NoStop}%
\bibitem [{\citenamefont {Wingstrand}\ \emph {et~al.}(2015)\citenamefont
  {Wingstrand}, \citenamefont {Alvarez}, \citenamefont {Huang},\ and\
  \citenamefont {Hassager}}]{wingstrand2015linear}%
  \BibitemOpen
  \bibfield  {author} {\bibinfo {author} {\bibfnamefont {S.~L.}\ \bibnamefont
  {Wingstrand}}, \bibinfo {author} {\bibfnamefont {N.~J.}\ \bibnamefont
  {Alvarez}}, \bibinfo {author} {\bibfnamefont {Q.}~\bibnamefont {Huang}}, \
  and\ \bibinfo {author} {\bibfnamefont {O.}~\bibnamefont {Hassager}},\
  }\href@noop {} {\bibfield  {journal} {\bibinfo  {journal} {Physical review
  letters}\ }\textbf {\bibinfo {volume} {115}},\ \bibinfo {pages} {078302}
  (\bibinfo {year} {2015})}\BibitemShut {NoStop}%
\bibitem [{\citenamefont {Graham}\ \emph {et~al.}(2003)\citenamefont {Graham},
  \citenamefont {Likhtman}, \citenamefont {McLeish},\ and\ \citenamefont
  {Milner}}]{graham2003microscopic}%
  \BibitemOpen
  \bibfield  {author} {\bibinfo {author} {\bibfnamefont {R.~S.}\ \bibnamefont
  {Graham}}, \bibinfo {author} {\bibfnamefont {A.~E.}\ \bibnamefont
  {Likhtman}}, \bibinfo {author} {\bibfnamefont {T.~C.}\ \bibnamefont
  {McLeish}}, \ and\ \bibinfo {author} {\bibfnamefont {S.~T.}\ \bibnamefont
  {Milner}},\ }\href@noop {} {\bibfield  {journal} {\bibinfo  {journal}
  {Journal of Rheology}\ }\textbf {\bibinfo {volume} {47}},\ \bibinfo {pages}
  {1171} (\bibinfo {year} {2003})}\BibitemShut {NoStop}%
\bibitem [{\citenamefont {Doi}\ and\ \citenamefont
  {Edwards}(1988)}]{doi1988theory}%
  \BibitemOpen
  \bibfield  {author} {\bibinfo {author} {\bibfnamefont {M.}~\bibnamefont
  {Doi}}\ and\ \bibinfo {author} {\bibfnamefont {S.~F.}\ \bibnamefont
  {Edwards}},\ }\href@noop {} {\emph {\bibinfo {title} {The theory of polymer
  dynamics}}},\ Vol.~\bibinfo {volume} {73}\ (\bibinfo  {publisher} {oxford
  university press},\ \bibinfo {year} {1988})\BibitemShut {NoStop}%
\bibitem [{\citenamefont {de~Gennes}(1971)}]{de1971reptation}%
  \BibitemOpen
  \bibfield  {author} {\bibinfo {author} {\bibfnamefont {P.-G.}\ \bibnamefont
  {de~Gennes}},\ }\href@noop {} {\bibfield  {journal} {\bibinfo  {journal} {The
  journal of chemical physics}\ }\textbf {\bibinfo {volume} {55}},\ \bibinfo
  {pages} {572} (\bibinfo {year} {1971})}\BibitemShut {NoStop}%
\bibitem [{\citenamefont {Han}(2007)}]{han2007rheology}%
  \BibitemOpen
  \bibfield  {author} {\bibinfo {author} {\bibfnamefont {C.~D.}\ \bibnamefont
  {Han}},\ }\href@noop {} {\emph {\bibinfo {title} {Rheology and Processing of
  Polymeric Materials: Volume 1: Polymer Rheology}}},\ Vol.~\bibinfo {volume}
  {1}\ (\bibinfo  {publisher} {Oxford University Press on Demand},\ \bibinfo
  {year} {2007})\BibitemShut {NoStop}%
\bibitem [{\citenamefont {Chennevi{\`e}re}\ \emph {et~al.}(2016)\citenamefont
  {Chennevi{\`e}re}, \citenamefont {Cousin}, \citenamefont {Bou{\'e}},
  \citenamefont {Drockenmuller}, \citenamefont {Shull}, \citenamefont
  {L{\'e}ger},\ and\ \citenamefont {Restagno}}]{chenneviere2016direct}%
  \BibitemOpen
  \bibfield  {author} {\bibinfo {author} {\bibfnamefont {A.}~\bibnamefont
  {Chennevi{\`e}re}}, \bibinfo {author} {\bibfnamefont {F.}~\bibnamefont
  {Cousin}}, \bibinfo {author} {\bibfnamefont {F.}~\bibnamefont {Bou{\'e}}},
  \bibinfo {author} {\bibfnamefont {E.}~\bibnamefont {Drockenmuller}}, \bibinfo
  {author} {\bibfnamefont {K.~R.}\ \bibnamefont {Shull}}, \bibinfo {author}
  {\bibfnamefont {L.}~\bibnamefont {L{\'e}ger}}, \ and\ \bibinfo {author}
  {\bibfnamefont {F.}~\bibnamefont {Restagno}},\ }\href@noop {} {\bibfield
  {journal} {\bibinfo  {journal} {Macromolecules}\ }\textbf {\bibinfo {volume}
  {49}},\ \bibinfo {pages} {2348} (\bibinfo {year} {2016})}\BibitemShut
  {NoStop}%
\bibitem [{\citenamefont {Korolkovas}\ \emph {et~al.}(2017)\citenamefont
  {Korolkovas}, \citenamefont {Rodriguez-Emmenegger}, \citenamefont {de~los
  Santos~Pereira}, \citenamefont {Chennevi{\`e}re}, \citenamefont {Restagno},
  \citenamefont {Wolff}, \citenamefont {Adlmann}, \citenamefont {Dennison},\
  and\ \citenamefont {Gutfreund}}]{korolkovas2017polymer}%
  \BibitemOpen
  \bibfield  {author} {\bibinfo {author} {\bibfnamefont {A.}~\bibnamefont
  {Korolkovas}}, \bibinfo {author} {\bibfnamefont {C.}~\bibnamefont
  {Rodriguez-Emmenegger}}, \bibinfo {author} {\bibfnamefont {A.}~\bibnamefont
  {de~los Santos~Pereira}}, \bibinfo {author} {\bibfnamefont {A.}~\bibnamefont
  {Chennevi{\`e}re}}, \bibinfo {author} {\bibfnamefont {F.}~\bibnamefont
  {Restagno}}, \bibinfo {author} {\bibfnamefont {M.}~\bibnamefont {Wolff}},
  \bibinfo {author} {\bibfnamefont {F.~A.}\ \bibnamefont {Adlmann}}, \bibinfo
  {author} {\bibfnamefont {A.~J.~C.}\ \bibnamefont {Dennison}}, \ and\ \bibinfo
  {author} {\bibfnamefont {P.}~\bibnamefont {Gutfreund}},\ }\href {\doibase
  10.1021/acs.macromol.6b02525} {\bibfield  {journal} {\bibinfo  {journal}
  {Macromolecules}\ }\textbf {\bibinfo {volume} {50}},\ \bibinfo {pages} {1215}
  (\bibinfo {year} {2017})}\BibitemShut {NoStop}%
\bibitem [{\citenamefont {Yearley}\ \emph {et~al.}(2010)\citenamefont
  {Yearley}, \citenamefont {Sasa}, \citenamefont {Welch}, \citenamefont
  {Taylor}, \citenamefont {Kupcho}, \citenamefont {Gilbertson},\ and\
  \citenamefont {Hjelm}}]{Yearley2010}%
  \BibitemOpen
  \bibfield  {author} {\bibinfo {author} {\bibfnamefont {E.~J.}\ \bibnamefont
  {Yearley}}, \bibinfo {author} {\bibfnamefont {L.~A.}\ \bibnamefont {Sasa}},
  \bibinfo {author} {\bibfnamefont {C.~F.}\ \bibnamefont {Welch}}, \bibinfo
  {author} {\bibfnamefont {M.~A.}\ \bibnamefont {Taylor}}, \bibinfo {author}
  {\bibfnamefont {K.~M.}\ \bibnamefont {Kupcho}}, \bibinfo {author}
  {\bibfnamefont {R.~D.}\ \bibnamefont {Gilbertson}}, \ and\ \bibinfo {author}
  {\bibfnamefont {R.~P.}\ \bibnamefont {Hjelm}},\ }\href {\doibase
  10.1063/1.3374121} {\bibfield  {journal} {\bibinfo  {journal} {Review of
  Scientific Instruments}\ }\textbf {\bibinfo {volume} {81}},\ \bibinfo {pages}
  {045109} (\bibinfo {year} {2010})},\ \Eprint
  {http://arxiv.org/abs/http://dx.doi.org/10.1063/1.3374121}
  {http://dx.doi.org/10.1063/1.3374121} \BibitemShut {NoStop}%
\bibitem [{\citenamefont {Bent}\ \emph {et~al.}(2003)\citenamefont {Bent},
  \citenamefont {Hutchings}, \citenamefont {Richards}, \citenamefont {Gough},
  \citenamefont {Spares}, \citenamefont {Coates}, \citenamefont {Grillo},
  \citenamefont {Harlen}, \citenamefont {Read}, \citenamefont {Graham},
  \citenamefont {Likhtman}, \citenamefont {Groves}, \citenamefont {Nicholson},\
  and\ \citenamefont {McLeish}}]{Bent2003}%
  \BibitemOpen
  \bibfield  {author} {\bibinfo {author} {\bibfnamefont {J.}~\bibnamefont
  {Bent}}, \bibinfo {author} {\bibfnamefont {L.~R.}\ \bibnamefont {Hutchings}},
  \bibinfo {author} {\bibfnamefont {R.~W.}\ \bibnamefont {Richards}}, \bibinfo
  {author} {\bibfnamefont {T.}~\bibnamefont {Gough}}, \bibinfo {author}
  {\bibfnamefont {R.}~\bibnamefont {Spares}}, \bibinfo {author} {\bibfnamefont
  {P.~D.}\ \bibnamefont {Coates}}, \bibinfo {author} {\bibfnamefont
  {I.}~\bibnamefont {Grillo}}, \bibinfo {author} {\bibfnamefont {O.~G.}\
  \bibnamefont {Harlen}}, \bibinfo {author} {\bibfnamefont {D.~J.}\
  \bibnamefont {Read}}, \bibinfo {author} {\bibfnamefont {R.~S.}\ \bibnamefont
  {Graham}}, \bibinfo {author} {\bibfnamefont {A.~E.}\ \bibnamefont
  {Likhtman}}, \bibinfo {author} {\bibfnamefont {D.~J.}\ \bibnamefont
  {Groves}}, \bibinfo {author} {\bibfnamefont {T.~M.}\ \bibnamefont
  {Nicholson}}, \ and\ \bibinfo {author} {\bibfnamefont {T.~C.~B.}\
  \bibnamefont {McLeish}},\ }\href {\doibase 10.1126/science.1086952}
  {\bibfield  {journal} {\bibinfo  {journal} {Science}\ }\textbf {\bibinfo
  {volume} {301}},\ \bibinfo {pages} {1691} (\bibinfo {year} {2003})},\ \Eprint
  {http://arxiv.org/abs/http://science.sciencemag.org/content/301/5640/1691.full.pdf}
  {http://science.sciencemag.org/content/301/5640/1691.full.pdf} \BibitemShut
  {NoStop}%
\bibitem [{\citenamefont {Graham}\ \emph {et~al.}(2006)\citenamefont {Graham},
  \citenamefont {Bent}, \citenamefont {Hutchings}, \citenamefont {Richards},
  \citenamefont {Groves}, \citenamefont {Embery}, \citenamefont {Nicholson},
  \citenamefont {McLeish}, \citenamefont {Likhtman}, \citenamefont {Harlen},
  \citenamefont {Read}, \citenamefont {Gough}, \citenamefont {Spares},
  \citenamefont {Coates},\ and\ \citenamefont {Grillo}}]{Graham2006}%
  \BibitemOpen
  \bibfield  {author} {\bibinfo {author} {\bibfnamefont {R.~S.}\ \bibnamefont
  {Graham}}, \bibinfo {author} {\bibfnamefont {J.}~\bibnamefont {Bent}},
  \bibinfo {author} {\bibfnamefont {L.~R.}\ \bibnamefont {Hutchings}}, \bibinfo
  {author} {\bibfnamefont {R.~W.}\ \bibnamefont {Richards}}, \bibinfo {author}
  {\bibfnamefont {D.~J.}\ \bibnamefont {Groves}}, \bibinfo {author}
  {\bibfnamefont {J.}~\bibnamefont {Embery}}, \bibinfo {author} {\bibfnamefont
  {T.~M.}\ \bibnamefont {Nicholson}}, \bibinfo {author} {\bibfnamefont
  {T.~C.~B.}\ \bibnamefont {McLeish}}, \bibinfo {author} {\bibfnamefont
  {A.~E.}\ \bibnamefont {Likhtman}}, \bibinfo {author} {\bibfnamefont {O.~G.}\
  \bibnamefont {Harlen}}, \bibinfo {author} {\bibfnamefont {D.~J.}\
  \bibnamefont {Read}}, \bibinfo {author} {\bibfnamefont {T.}~\bibnamefont
  {Gough}}, \bibinfo {author} {\bibfnamefont {R.}~\bibnamefont {Spares}},
  \bibinfo {author} {\bibfnamefont {P.~D.}\ \bibnamefont {Coates}}, \ and\
  \bibinfo {author} {\bibfnamefont {I.}~\bibnamefont {Grillo}},\ }\href
  {\doibase 10.1021/ma052357z} {\bibfield  {journal} {\bibinfo  {journal}
  {Macromolecules}\ }\textbf {\bibinfo {volume} {39}},\ \bibinfo {pages} {2700}
  (\bibinfo {year} {2006})},\ \Eprint
  {http://arxiv.org/abs/http://dx.doi.org/10.1021/ma052357z}
  {http://dx.doi.org/10.1021/ma052357z} \BibitemShut {NoStop}%
\bibitem [{\citenamefont {Clarke}\ \emph {et~al.}(2010)\citenamefont {Clarke},
  \citenamefont {De~Luca}, \citenamefont {Buxton}, \citenamefont {Hutchings},
  \citenamefont {Gough}, \citenamefont {Grillo}, \citenamefont {Graham},
  \citenamefont {Jagannathan}, \citenamefont {Klein},\ and\ \citenamefont
  {McLeish}}]{Clarke2010}%
  \BibitemOpen
  \bibfield  {author} {\bibinfo {author} {\bibfnamefont {N.}~\bibnamefont
  {Clarke}}, \bibinfo {author} {\bibfnamefont {E.}~\bibnamefont {De~Luca}},
  \bibinfo {author} {\bibfnamefont {G.}~\bibnamefont {Buxton}}, \bibinfo
  {author} {\bibfnamefont {L.~R.}\ \bibnamefont {Hutchings}}, \bibinfo {author}
  {\bibfnamefont {T.}~\bibnamefont {Gough}}, \bibinfo {author} {\bibfnamefont
  {I.}~\bibnamefont {Grillo}}, \bibinfo {author} {\bibfnamefont {R.~S.}\
  \bibnamefont {Graham}}, \bibinfo {author} {\bibfnamefont {K.}~\bibnamefont
  {Jagannathan}}, \bibinfo {author} {\bibfnamefont {D.~H.}\ \bibnamefont
  {Klein}}, \ and\ \bibinfo {author} {\bibfnamefont {T.~C.~B.}\ \bibnamefont
  {McLeish}},\ }\href {\doibase 10.1021/ma902324f} {\bibfield  {journal}
  {\bibinfo  {journal} {Macromolecules}\ }\textbf {\bibinfo {volume} {43}},\
  \bibinfo {pages} {1539} (\bibinfo {year} {2010})},\ \Eprint
  {http://arxiv.org/abs/http://dx.doi.org/10.1021/ma902324f}
  {http://dx.doi.org/10.1021/ma902324f} \BibitemShut {NoStop}%
\bibitem [{\citenamefont {Muller}, \citenamefont {Pesce},\ and\ \citenamefont
  {Picot}(1993)}]{Muller1993}%
  \BibitemOpen
  \bibfield  {author} {\bibinfo {author} {\bibfnamefont {R.}~\bibnamefont
  {Muller}}, \bibinfo {author} {\bibfnamefont {J.~J.}\ \bibnamefont {Pesce}}, \
  and\ \bibinfo {author} {\bibfnamefont {C.}~\bibnamefont {Picot}},\ }\href
  {\doibase 10.1021/ma00068a044} {\bibfield  {journal} {\bibinfo  {journal}
  {Macromolecules}\ }\textbf {\bibinfo {volume} {26}},\ \bibinfo {pages} {4356}
  (\bibinfo {year} {1993})},\ \Eprint
  {http://arxiv.org/abs/http://dx.doi.org/10.1021/ma00068a044}
  {http://dx.doi.org/10.1021/ma00068a044} \BibitemShut {NoStop}%
\bibitem [{\citenamefont {Blanchard}\ \emph {et~al.}(2005)\citenamefont
  {Blanchard}, \citenamefont {Graham}, \citenamefont {Heinrich}, \citenamefont
  {Pyckhout-Hintzen}, \citenamefont {Richter}, \citenamefont {Likhtman},
  \citenamefont {McLeish}, \citenamefont {Read}, \citenamefont {Straube},\ and\
  \citenamefont {Kohlbrecher}}]{Blanchard2005}%
  \BibitemOpen
  \bibfield  {author} {\bibinfo {author} {\bibfnamefont {A.}~\bibnamefont
  {Blanchard}}, \bibinfo {author} {\bibfnamefont {R.~S.}\ \bibnamefont
  {Graham}}, \bibinfo {author} {\bibfnamefont {M.}~\bibnamefont {Heinrich}},
  \bibinfo {author} {\bibfnamefont {W.}~\bibnamefont {Pyckhout-Hintzen}},
  \bibinfo {author} {\bibfnamefont {D.}~\bibnamefont {Richter}}, \bibinfo
  {author} {\bibfnamefont {A.~E.}\ \bibnamefont {Likhtman}}, \bibinfo {author}
  {\bibfnamefont {T.~C.~B.}\ \bibnamefont {McLeish}}, \bibinfo {author}
  {\bibfnamefont {D.~J.}\ \bibnamefont {Read}}, \bibinfo {author}
  {\bibfnamefont {E.}~\bibnamefont {Straube}}, \ and\ \bibinfo {author}
  {\bibfnamefont {J.}~\bibnamefont {Kohlbrecher}},\ }\href {\doibase
  10.1103/PhysRevLett.95.166001} {\bibfield  {journal} {\bibinfo  {journal}
  {Physical Review Letters}\ }\textbf {\bibinfo {volume} {95}},\ \bibinfo
  {pages} {166001} (\bibinfo {year} {2005})}\BibitemShut {NoStop}%
\bibitem [{\citenamefont {Bou{\'e}}(1987)}]{Boue1987}%
  \BibitemOpen
  \bibfield  {author} {\bibinfo {author} {\bibfnamefont {F.}~\bibnamefont
  {Bou{\'e}}},\ }\enquote {\bibinfo {title} {Transient relaxation mechanisms in
  elongated melts and rubbers investigated by small angle neutron
  scattering},}\ in\ \href {\doibase 10.1007/BFb0024042} {\emph {\bibinfo
  {booktitle} {Polymer Physics}}}\ (\bibinfo  {publisher} {Springer Berlin
  Heidelberg},\ \bibinfo {address} {Berlin, Heidelberg},\ \bibinfo {year}
  {1987})\ pp.\ \bibinfo {pages} {47--101}\BibitemShut {NoStop}%
\bibitem [{\citenamefont {Kirkensgaard}\ \emph {et~al.}(2016)\citenamefont
  {Kirkensgaard}, \citenamefont {Hengeller}, \citenamefont {Dorokhin},
  \citenamefont {Huang}, \citenamefont {Garvey}, \citenamefont {Almdal},
  \citenamefont {Hassager},\ and\ \citenamefont
  {Mortensen}}]{Kirkensgaard2016}%
  \BibitemOpen
  \bibfield  {author} {\bibinfo {author} {\bibfnamefont {J.~J.~K.}\
  \bibnamefont {Kirkensgaard}}, \bibinfo {author} {\bibfnamefont
  {L.}~\bibnamefont {Hengeller}}, \bibinfo {author} {\bibfnamefont
  {A.}~\bibnamefont {Dorokhin}}, \bibinfo {author} {\bibfnamefont
  {Q.}~\bibnamefont {Huang}}, \bibinfo {author} {\bibfnamefont {C.~J.}\
  \bibnamefont {Garvey}}, \bibinfo {author} {\bibfnamefont {K.}~\bibnamefont
  {Almdal}}, \bibinfo {author} {\bibfnamefont {O.}~\bibnamefont {Hassager}}, \
  and\ \bibinfo {author} {\bibfnamefont {K.}~\bibnamefont {Mortensen}},\
  }\href@noop {} {\bibfield  {journal} {\bibinfo  {journal} {Physical Review
  E}\ }\textbf {\bibinfo {volume} {94}},\ \bibinfo {pages} {020502} (\bibinfo
  {year} {2016})}\BibitemShut {NoStop}%
\bibitem [{\citenamefont {Wang}\ \emph {et~al.}(2017)\citenamefont {Wang},
  \citenamefont {Lam}, \citenamefont {Chen}, \citenamefont {Wang},
  \citenamefont {Liu}, \citenamefont {Liu}, \citenamefont {Porcar},
  \citenamefont {Stanley}, \citenamefont {Zhao}, \citenamefont {Hong} \emph
  {et~al.}}]{wang2017fingerprinting}%
  \BibitemOpen
  \bibfield  {author} {\bibinfo {author} {\bibfnamefont {Z.}~\bibnamefont
  {Wang}}, \bibinfo {author} {\bibfnamefont {C.~N.}\ \bibnamefont {Lam}},
  \bibinfo {author} {\bibfnamefont {W.-R.}\ \bibnamefont {Chen}}, \bibinfo
  {author} {\bibfnamefont {W.}~\bibnamefont {Wang}}, \bibinfo {author}
  {\bibfnamefont {J.}~\bibnamefont {Liu}}, \bibinfo {author} {\bibfnamefont
  {Y.}~\bibnamefont {Liu}}, \bibinfo {author} {\bibfnamefont {L.}~\bibnamefont
  {Porcar}}, \bibinfo {author} {\bibfnamefont {C.~B.}\ \bibnamefont {Stanley}},
  \bibinfo {author} {\bibfnamefont {Z.}~\bibnamefont {Zhao}}, \bibinfo {author}
  {\bibfnamefont {K.}~\bibnamefont {Hong}},  \emph {et~al.},\ }\href@noop {}
  {\bibfield  {journal} {\bibinfo  {journal} {Physical Review X}\ }\textbf
  {\bibinfo {volume} {7}},\ \bibinfo {pages} {031003} (\bibinfo {year}
  {2017})}\BibitemShut {NoStop}%
\bibitem [{\citenamefont {Goo{\ss}en}\ \emph {et~al.}(2015)\citenamefont
  {Goo{\ss}en}, \citenamefont {Krutyeva}, \citenamefont {Sharp}, \citenamefont
  {Feoktystov}, \citenamefont {Allgaier}, \citenamefont {Pyckhout-Hintzen},
  \citenamefont {Wischnewski},\ and\ \citenamefont
  {Richter}}]{goossen2015sensing}%
  \BibitemOpen
  \bibfield  {author} {\bibinfo {author} {\bibfnamefont {S.}~\bibnamefont
  {Goo{\ss}en}}, \bibinfo {author} {\bibfnamefont {M.}~\bibnamefont
  {Krutyeva}}, \bibinfo {author} {\bibfnamefont {M.}~\bibnamefont {Sharp}},
  \bibinfo {author} {\bibfnamefont {A.}~\bibnamefont {Feoktystov}}, \bibinfo
  {author} {\bibfnamefont {J.}~\bibnamefont {Allgaier}}, \bibinfo {author}
  {\bibfnamefont {W.}~\bibnamefont {Pyckhout-Hintzen}}, \bibinfo {author}
  {\bibfnamefont {A.}~\bibnamefont {Wischnewski}}, \ and\ \bibinfo {author}
  {\bibfnamefont {D.}~\bibnamefont {Richter}},\ }\href@noop {} {\bibfield
  {journal} {\bibinfo  {journal} {Physical review letters}\ }\textbf {\bibinfo
  {volume} {115}},\ \bibinfo {pages} {148302} (\bibinfo {year}
  {2015})}\BibitemShut {NoStop}%
\bibitem [{\citenamefont {Schleger}\ \emph {et~al.}(1998)\citenamefont
  {Schleger}, \citenamefont {Farago}, \citenamefont {Lartigue}, \citenamefont
  {Kollmar},\ and\ \citenamefont {Richter}}]{schleger1998clear}%
  \BibitemOpen
  \bibfield  {author} {\bibinfo {author} {\bibfnamefont {P.}~\bibnamefont
  {Schleger}}, \bibinfo {author} {\bibfnamefont {B.}~\bibnamefont {Farago}},
  \bibinfo {author} {\bibfnamefont {C.}~\bibnamefont {Lartigue}}, \bibinfo
  {author} {\bibfnamefont {A.}~\bibnamefont {Kollmar}}, \ and\ \bibinfo
  {author} {\bibfnamefont {D.}~\bibnamefont {Richter}},\ }\href@noop {}
  {\bibfield  {journal} {\bibinfo  {journal} {Physical review letters}\
  }\textbf {\bibinfo {volume} {81}},\ \bibinfo {pages} {124} (\bibinfo {year}
  {1998})}\BibitemShut {NoStop}%
\bibitem [{\citenamefont {Wischnewski}\ \emph {et~al.}(2002)\citenamefont
  {Wischnewski}, \citenamefont {Monkenbusch}, \citenamefont {Willner},
  \citenamefont {Richter}, \citenamefont {Likhtman}, \citenamefont {McLeish},\
  and\ \citenamefont {Farago}}]{wischnewski2002molecular}%
  \BibitemOpen
  \bibfield  {author} {\bibinfo {author} {\bibfnamefont {A.}~\bibnamefont
  {Wischnewski}}, \bibinfo {author} {\bibfnamefont {M.}~\bibnamefont
  {Monkenbusch}}, \bibinfo {author} {\bibfnamefont {L.}~\bibnamefont
  {Willner}}, \bibinfo {author} {\bibfnamefont {D.}~\bibnamefont {Richter}},
  \bibinfo {author} {\bibfnamefont {A.}~\bibnamefont {Likhtman}}, \bibinfo
  {author} {\bibfnamefont {T.}~\bibnamefont {McLeish}}, \ and\ \bibinfo
  {author} {\bibfnamefont {B.}~\bibnamefont {Farago}},\ }\href@noop {}
  {\bibfield  {journal} {\bibinfo  {journal} {Physical review letters}\
  }\textbf {\bibinfo {volume} {88}},\ \bibinfo {pages} {058301} (\bibinfo
  {year} {2002})}\BibitemShut {NoStop}%
\bibitem [{\citenamefont {Richter}\ \emph
  {et~al.}(1993{\natexlab{a}})\citenamefont {Richter}, \citenamefont {Ewen},
  \citenamefont {Fetters}, \citenamefont {Huang},\ and\ \citenamefont
  {Farago}}]{richter1993dynamics}%
  \BibitemOpen
  \bibfield  {author} {\bibinfo {author} {\bibfnamefont {D.}~\bibnamefont
  {Richter}}, \bibinfo {author} {\bibfnamefont {B.}~\bibnamefont {Ewen}},
  \bibinfo {author} {\bibfnamefont {L.}~\bibnamefont {Fetters}}, \bibinfo
  {author} {\bibfnamefont {J.}~\bibnamefont {Huang}}, \ and\ \bibinfo {author}
  {\bibfnamefont {B.}~\bibnamefont {Farago}},\ }in\ \href@noop {} {\emph
  {\bibinfo {booktitle} {Application of Scattering Methods to the Dynamics of
  Polymer Systems}}}\ (\bibinfo  {publisher} {Springer},\ \bibinfo {year}
  {1993})\ pp.\ \bibinfo {pages} {130--134}\BibitemShut {NoStop}%
\bibitem [{\citenamefont {Richter}\ \emph
  {et~al.}(1993{\natexlab{b}})\citenamefont {Richter}, \citenamefont {Farago},
  \citenamefont {Butera}, \citenamefont {Fetters}, \citenamefont {Huang},\ and\
  \citenamefont {Ewen}}]{richter1993origins}%
  \BibitemOpen
  \bibfield  {author} {\bibinfo {author} {\bibfnamefont {D.}~\bibnamefont
  {Richter}}, \bibinfo {author} {\bibfnamefont {B.}~\bibnamefont {Farago}},
  \bibinfo {author} {\bibfnamefont {R.}~\bibnamefont {Butera}}, \bibinfo
  {author} {\bibfnamefont {L.}~\bibnamefont {Fetters}}, \bibinfo {author}
  {\bibfnamefont {J.}~\bibnamefont {Huang}}, \ and\ \bibinfo {author}
  {\bibfnamefont {B.}~\bibnamefont {Ewen}},\ }\href@noop {} {\bibfield
  {journal} {\bibinfo  {journal} {Macromolecules}\ }\textbf {\bibinfo {volume}
  {26}},\ \bibinfo {pages} {795} (\bibinfo {year}
  {1993}{\natexlab{b}})}\BibitemShut {NoStop}%
\bibitem [{\citenamefont {Pyckhout-Hintzen}, \citenamefont {Wischnewski},\ and\
  \citenamefont {Richter}(2016)}]{pyckhout2016mixtures}%
  \BibitemOpen
  \bibfield  {author} {\bibinfo {author} {\bibfnamefont {W.}~\bibnamefont
  {Pyckhout-Hintzen}}, \bibinfo {author} {\bibfnamefont {A.}~\bibnamefont
  {Wischnewski}}, \ and\ \bibinfo {author} {\bibfnamefont {D.}~\bibnamefont
  {Richter}},\ }\href@noop {} {\bibfield  {journal} {\bibinfo  {journal}
  {Polymer}\ }\textbf {\bibinfo {volume} {105}},\ \bibinfo {pages} {378}
  (\bibinfo {year} {2016})}\BibitemShut {NoStop}%
\bibitem [{\citenamefont {Rao}(2014)}]{rao2014rheology}%
  \BibitemOpen
  \bibfield  {author} {\bibinfo {author} {\bibfnamefont {M.~A.}\ \bibnamefont
  {Rao}},\ }in\ \href@noop {} {\emph {\bibinfo {booktitle} {Rheology of fluid,
  semisolid, and solid foods}}}\ (\bibinfo  {publisher} {Springer},\ \bibinfo
  {year} {2014})\ pp.\ \bibinfo {pages} {161--229}\BibitemShut {NoStop}%
\bibitem [{\citenamefont {Chhabra}\ and\ \citenamefont
  {Richardson}(2011)}]{chhabra2011non}%
  \BibitemOpen
  \bibfield  {author} {\bibinfo {author} {\bibfnamefont {R.~P.}\ \bibnamefont
  {Chhabra}}\ and\ \bibinfo {author} {\bibfnamefont {J.~F.}\ \bibnamefont
  {Richardson}},\ }\href@noop {} {\emph {\bibinfo {title} {Non-Newtonian flow
  and applied rheology: engineering applications}}}\ (\bibinfo  {publisher}
  {Butterworth-Heinemann},\ \bibinfo {year} {2011})\BibitemShut {NoStop}%
\bibitem [{\citenamefont {Bartczak}(2005)}]{bartczak2005effect}%
  \BibitemOpen
  \bibfield  {author} {\bibinfo {author} {\bibfnamefont {Z.}~\bibnamefont
  {Bartczak}},\ }\href@noop {} {\bibfield  {journal} {\bibinfo  {journal}
  {Macromolecules}\ }\textbf {\bibinfo {volume} {38}},\ \bibinfo {pages} {7702}
  (\bibinfo {year} {2005})}\BibitemShut {NoStop}%
\bibitem [{\citenamefont {Bartczak}(2010)}]{bartczak2010effect}%
  \BibitemOpen
  \bibfield  {author} {\bibinfo {author} {\bibfnamefont {Z.}~\bibnamefont
  {Bartczak}},\ }\href@noop {} {\bibfield  {journal} {\bibinfo  {journal}
  {Journal of Polymer Science Part B: Polymer Physics}\ }\textbf {\bibinfo
  {volume} {48}},\ \bibinfo {pages} {276} (\bibinfo {year} {2010})}\BibitemShut
  {NoStop}%
\bibitem [{\citenamefont {Shenoy}\ \emph {et~al.}(2005)\citenamefont {Shenoy},
  \citenamefont {Bates}, \citenamefont {Frisch},\ and\ \citenamefont
  {Wnek}}]{shenoy2005role}%
  \BibitemOpen
  \bibfield  {author} {\bibinfo {author} {\bibfnamefont {S.~L.}\ \bibnamefont
  {Shenoy}}, \bibinfo {author} {\bibfnamefont {W.~D.}\ \bibnamefont {Bates}},
  \bibinfo {author} {\bibfnamefont {H.~L.}\ \bibnamefont {Frisch}}, \ and\
  \bibinfo {author} {\bibfnamefont {G.~E.}\ \bibnamefont {Wnek}},\ }\href@noop
  {} {\bibfield  {journal} {\bibinfo  {journal} {Polymer}\ }\textbf {\bibinfo
  {volume} {46}},\ \bibinfo {pages} {3372} (\bibinfo {year}
  {2005})}\BibitemShut {NoStop}%
\bibitem [{\citenamefont {Wever}, \citenamefont {Picchioni},\ and\
  \citenamefont {Broekhuis}(2011)}]{wever2011polymers}%
  \BibitemOpen
  \bibfield  {author} {\bibinfo {author} {\bibfnamefont {D.}~\bibnamefont
  {Wever}}, \bibinfo {author} {\bibfnamefont {F.}~\bibnamefont {Picchioni}}, \
  and\ \bibinfo {author} {\bibfnamefont {A.}~\bibnamefont {Broekhuis}},\
  }\href@noop {} {\bibfield  {journal} {\bibinfo  {journal} {Progress in
  Polymer Science}\ }\textbf {\bibinfo {volume} {36}},\ \bibinfo {pages} {1558}
  (\bibinfo {year} {2011})}\BibitemShut {NoStop}%
\bibitem [{\citenamefont {Tsang}\ \emph {et~al.}(2017)\citenamefont {Tsang},
  \citenamefont {Dell}, \citenamefont {Jiang}, \citenamefont {Schweizer},\ and\
  \citenamefont {Granick}}]{tsang2017dynamic}%
  \BibitemOpen
  \bibfield  {author} {\bibinfo {author} {\bibfnamefont {B.}~\bibnamefont
  {Tsang}}, \bibinfo {author} {\bibfnamefont {Z.~E.}\ \bibnamefont {Dell}},
  \bibinfo {author} {\bibfnamefont {L.}~\bibnamefont {Jiang}}, \bibinfo
  {author} {\bibfnamefont {K.~S.}\ \bibnamefont {Schweizer}}, \ and\ \bibinfo
  {author} {\bibfnamefont {S.}~\bibnamefont {Granick}},\ }\href@noop {}
  {\bibfield  {journal} {\bibinfo  {journal} {Proceedings of the National
  Academy of Sciences}\ ,\ \bibinfo {pages} {201620935}} (\bibinfo {year}
  {2017})}\BibitemShut {NoStop}%
\bibitem [{\citenamefont {June}\ \emph {et~al.}(2009)\citenamefont {June},
  \citenamefont {Mejia}, \citenamefont {Barone},\ and\ \citenamefont
  {Fyhrie}}]{june2009cartilage}%
  \BibitemOpen
  \bibfield  {author} {\bibinfo {author} {\bibfnamefont {R.~K.}\ \bibnamefont
  {June}}, \bibinfo {author} {\bibfnamefont {K.~L.}\ \bibnamefont {Mejia}},
  \bibinfo {author} {\bibfnamefont {J.~R.}\ \bibnamefont {Barone}}, \ and\
  \bibinfo {author} {\bibfnamefont {D.~P.}\ \bibnamefont {Fyhrie}},\
  }\href@noop {} {\bibfield  {journal} {\bibinfo  {journal} {Osteoarthritis and
  Cartilage}\ }\textbf {\bibinfo {volume} {17}},\ \bibinfo {pages} {669}
  (\bibinfo {year} {2009})}\BibitemShut {NoStop}%
\bibitem [{\citenamefont {Teixeira}\ \emph {et~al.}(2007)\citenamefont
  {Teixeira}, \citenamefont {Dambal}, \citenamefont {Richter}, \citenamefont
  {Shaqfeh},\ and\ \citenamefont {Chu}}]{teixeira2007individualistic}%
  \BibitemOpen
  \bibfield  {author} {\bibinfo {author} {\bibfnamefont {R.~E.}\ \bibnamefont
  {Teixeira}}, \bibinfo {author} {\bibfnamefont {A.~K.}\ \bibnamefont
  {Dambal}}, \bibinfo {author} {\bibfnamefont {D.~H.}\ \bibnamefont {Richter}},
  \bibinfo {author} {\bibfnamefont {E.~S.}\ \bibnamefont {Shaqfeh}}, \ and\
  \bibinfo {author} {\bibfnamefont {S.}~\bibnamefont {Chu}},\ }\href@noop {}
  {\bibfield  {journal} {\bibinfo  {journal} {Macromolecules}\ }\textbf
  {\bibinfo {volume} {40}},\ \bibinfo {pages} {2461} (\bibinfo {year}
  {2007})}\BibitemShut {NoStop}%
\bibitem [{\citenamefont {Vermant}\ \emph {et~al.}(1998)\citenamefont
  {Vermant}, \citenamefont {Walker}, \citenamefont {Moldenaers},\ and\
  \citenamefont {Mewis}}]{vermant1998orthogonal}%
  \BibitemOpen
  \bibfield  {author} {\bibinfo {author} {\bibfnamefont {J.}~\bibnamefont
  {Vermant}}, \bibinfo {author} {\bibfnamefont {L.}~\bibnamefont {Walker}},
  \bibinfo {author} {\bibfnamefont {P.}~\bibnamefont {Moldenaers}}, \ and\
  \bibinfo {author} {\bibfnamefont {J.}~\bibnamefont {Mewis}},\ }\href@noop {}
  {\bibfield  {journal} {\bibinfo  {journal} {Journal of non-newtonian fluid
  mechanics}\ }\textbf {\bibinfo {volume} {79}},\ \bibinfo {pages} {173}
  (\bibinfo {year} {1998})}\BibitemShut {NoStop}%
\bibitem [{\citenamefont {Watanabe}, \citenamefont {Matsumiya},\ and\
  \citenamefont {Inoue}(2003)}]{watanabe2003rheo}%
  \BibitemOpen
  \bibfield  {author} {\bibinfo {author} {\bibfnamefont {H.}~\bibnamefont
  {Watanabe}}, \bibinfo {author} {\bibfnamefont {Y.}~\bibnamefont {Matsumiya}},
  \ and\ \bibinfo {author} {\bibfnamefont {T.}~\bibnamefont {Inoue}},\
  }\href@noop {} {\bibfield  {journal} {\bibinfo  {journal} {Journal of
  Physics: Condensed Matter}\ }\textbf {\bibinfo {volume} {15}},\ \bibinfo
  {pages} {S909} (\bibinfo {year} {2003})}\BibitemShut {NoStop}%
\bibitem [{\citenamefont {B{\"o}hme}\ and\ \citenamefont
  {Scheler}(2011)}]{bohme2011nmr}%
  \BibitemOpen
  \bibfield  {author} {\bibinfo {author} {\bibfnamefont {U.}~\bibnamefont
  {B{\"o}hme}}\ and\ \bibinfo {author} {\bibfnamefont {U.}~\bibnamefont
  {Scheler}},\ }in\ \href@noop {} {\emph {\bibinfo {booktitle} {NMR
  Spectroscopy of Polymers: Innovative Strategies for Complex
  Macromolecules}}}\ (\bibinfo  {publisher} {ACS Publications},\ \bibinfo
  {year} {2011})\ pp.\ \bibinfo {pages} {431--438}\BibitemShut {NoStop}%
\bibitem [{\citenamefont {Kawecki}\ \emph {et~al.}(2017)\citenamefont
  {Kawecki}, \citenamefont {Adlmann}, \citenamefont {Gutfreund}, \citenamefont
  {Gupta}, \citenamefont {Falus}, \citenamefont {Zolnierczuk}, \citenamefont
  {Farago}, \citenamefont {Uhrig}, \citenamefont {Cochran},\ and\ \citenamefont
  {Wolff}}]{Kawecki2017}%
  \BibitemOpen
  \bibfield  {author} {\bibinfo {author} {\bibfnamefont {M.}~\bibnamefont
  {Kawecki}}, \bibinfo {author} {\bibfnamefont {F.~A.}\ \bibnamefont
  {Adlmann}}, \bibinfo {author} {\bibfnamefont {P.}~\bibnamefont {Gutfreund}},
  \bibinfo {author} {\bibfnamefont {S.}~\bibnamefont {Gupta}}, \bibinfo
  {author} {\bibfnamefont {P.}~\bibnamefont {Falus}}, \bibinfo {author}
  {\bibfnamefont {P.}~\bibnamefont {Zolnierczuk}}, \bibinfo {author}
  {\bibfnamefont {B.}~\bibnamefont {Farago}}, \bibinfo {author} {\bibfnamefont
  {D.}~\bibnamefont {Uhrig}}, \bibinfo {author} {\bibfnamefont
  {M.}~\bibnamefont {Cochran}}, \ and\ \bibinfo {author} {\bibfnamefont
  {M.}~\bibnamefont {Wolff}},\ }\href@noop {} {\enquote {\bibinfo {title}
  {Topological interactions in polymers under shear},}\ } (\bibinfo {year}
  {2017}),\ \bibinfo {note} {scientific Reports (under review)}\BibitemShut
  {NoStop}%
\bibitem [{\citenamefont {Aoyagi}\ and\ \citenamefont
  {Doi}(2000)}]{aoyagi2000molecular}%
  \BibitemOpen
  \bibfield  {author} {\bibinfo {author} {\bibfnamefont {T.}~\bibnamefont
  {Aoyagi}}\ and\ \bibinfo {author} {\bibfnamefont {M.}~\bibnamefont {Doi}},\
  }\href@noop {} {\bibfield  {journal} {\bibinfo  {journal} {Computational and
  Theoretical Polymer Science}\ }\textbf {\bibinfo {volume} {10}},\ \bibinfo
  {pages} {317} (\bibinfo {year} {2000})}\BibitemShut {NoStop}%
\bibitem [{\citenamefont {Kr{\"o}ger}\ and\ \citenamefont
  {Hess}(2000)}]{kroger2000rheological}%
  \BibitemOpen
  \bibfield  {author} {\bibinfo {author} {\bibfnamefont {M.}~\bibnamefont
  {Kr{\"o}ger}}\ and\ \bibinfo {author} {\bibfnamefont {S.}~\bibnamefont
  {Hess}},\ }\href@noop {} {\bibfield  {journal} {\bibinfo  {journal} {Physical
  review letters}\ }\textbf {\bibinfo {volume} {85}},\ \bibinfo {pages} {1128}
  (\bibinfo {year} {2000})}\BibitemShut {NoStop}%
\bibitem [{\citenamefont {Padding}\ and\ \citenamefont
  {Briels}(2003)}]{padding2003coarse}%
  \BibitemOpen
  \bibfield  {author} {\bibinfo {author} {\bibfnamefont {J.}~\bibnamefont
  {Padding}}\ and\ \bibinfo {author} {\bibfnamefont {W.}~\bibnamefont
  {Briels}},\ }\href@noop {} {\bibfield  {journal} {\bibinfo  {journal} {The
  Journal of chemical physics}\ }\textbf {\bibinfo {volume} {118}},\ \bibinfo
  {pages} {10276} (\bibinfo {year} {2003})}\BibitemShut {NoStop}%
\bibitem [{\citenamefont {Baig}, \citenamefont {Mavrantzas},\ and\
  \citenamefont {Kr{\"o}ger}(2010)}]{baig2010flow}%
  \BibitemOpen
  \bibfield  {author} {\bibinfo {author} {\bibfnamefont {C.}~\bibnamefont
  {Baig}}, \bibinfo {author} {\bibfnamefont {V.~G.}\ \bibnamefont
  {Mavrantzas}}, \ and\ \bibinfo {author} {\bibfnamefont {M.}~\bibnamefont
  {Kr{\"o}ger}},\ }\href@noop {} {\bibfield  {journal} {\bibinfo  {journal}
  {Macromolecules}\ }\textbf {\bibinfo {volume} {43}},\ \bibinfo {pages} {6886}
  (\bibinfo {year} {2010})}\BibitemShut {NoStop}%
\bibitem [{\citenamefont {Dorgan}, \citenamefont {Rorrer},\ and\ \citenamefont
  {Maupin}(2012)}]{dorgan2012parameter}%
  \BibitemOpen
  \bibfield  {author} {\bibinfo {author} {\bibfnamefont {J.~R.}\ \bibnamefont
  {Dorgan}}, \bibinfo {author} {\bibfnamefont {N.~A.}\ \bibnamefont {Rorrer}},
  \ and\ \bibinfo {author} {\bibfnamefont {C.~M.}\ \bibnamefont {Maupin}},\
  }\href@noop {} {\bibfield  {journal} {\bibinfo  {journal} {Macromolecules}\
  }\textbf {\bibinfo {volume} {45}},\ \bibinfo {pages} {8833} (\bibinfo {year}
  {2012})}\BibitemShut {NoStop}%
\bibitem [{\citenamefont {Halverson}\ \emph {et~al.}(2012)\citenamefont
  {Halverson}, \citenamefont {Grest}, \citenamefont {Grosberg},\ and\
  \citenamefont {Kremer}}]{halverson2012rheology}%
  \BibitemOpen
  \bibfield  {author} {\bibinfo {author} {\bibfnamefont {J.~D.}\ \bibnamefont
  {Halverson}}, \bibinfo {author} {\bibfnamefont {G.~S.}\ \bibnamefont
  {Grest}}, \bibinfo {author} {\bibfnamefont {A.~Y.}\ \bibnamefont {Grosberg}},
  \ and\ \bibinfo {author} {\bibfnamefont {K.}~\bibnamefont {Kremer}},\
  }\href@noop {} {\bibfield  {journal} {\bibinfo  {journal} {Physical review
  letters}\ }\textbf {\bibinfo {volume} {108}},\ \bibinfo {pages} {038301}
  (\bibinfo {year} {2012})}\BibitemShut {NoStop}%
\bibitem [{\citenamefont {Xu}, \citenamefont {Chen},\ and\ \citenamefont
  {An}(2014)}]{xu2014shear}%
  \BibitemOpen
  \bibfield  {author} {\bibinfo {author} {\bibfnamefont {X.}~\bibnamefont
  {Xu}}, \bibinfo {author} {\bibfnamefont {J.}~\bibnamefont {Chen}}, \ and\
  \bibinfo {author} {\bibfnamefont {L.}~\bibnamefont {An}},\ }\href@noop {}
  {\bibfield  {journal} {\bibinfo  {journal} {The Journal of chemical physics}\
  }\textbf {\bibinfo {volume} {140}},\ \bibinfo {pages} {174902} (\bibinfo
  {year} {2014})}\BibitemShut {NoStop}%
\bibitem [{\citenamefont {Kr{\"o}ger}, \citenamefont {Luap},\ and\
  \citenamefont {Muller}(1997)}]{kroger1997polymer}%
  \BibitemOpen
  \bibfield  {author} {\bibinfo {author} {\bibfnamefont {M.}~\bibnamefont
  {Kr{\"o}ger}}, \bibinfo {author} {\bibfnamefont {C.}~\bibnamefont {Luap}}, \
  and\ \bibinfo {author} {\bibfnamefont {R.}~\bibnamefont {Muller}},\
  }\href@noop {} {\bibfield  {journal} {\bibinfo  {journal} {Macromolecules}\
  }\textbf {\bibinfo {volume} {30}},\ \bibinfo {pages} {526} (\bibinfo {year}
  {1997})}\BibitemShut {NoStop}%
\bibitem [{\citenamefont {Cao}\ and\ \citenamefont
  {Likhtman}(2015)}]{cao2015simulating}%
  \BibitemOpen
  \bibfield  {author} {\bibinfo {author} {\bibfnamefont {J.}~\bibnamefont
  {Cao}}\ and\ \bibinfo {author} {\bibfnamefont {A.~E.}\ \bibnamefont
  {Likhtman}},\ }\href@noop {} {\bibfield  {journal} {\bibinfo  {journal} {ACS
  Macro Letters}\ }\textbf {\bibinfo {volume} {4}},\ \bibinfo {pages} {1376}
  (\bibinfo {year} {2015})}\BibitemShut {NoStop}%
\bibitem [{\citenamefont {Xu}\ \emph {et~al.}(2018)\citenamefont {Xu},
  \citenamefont {Carrillo}, \citenamefont {Lam}, \citenamefont {Sumpter},\ and\
  \citenamefont {Wang}}]{xu2018molecular}%
  \BibitemOpen
  \bibfield  {author} {\bibinfo {author} {\bibfnamefont {W.-S.}\ \bibnamefont
  {Xu}}, \bibinfo {author} {\bibfnamefont {J.-M.~Y.}\ \bibnamefont {Carrillo}},
  \bibinfo {author} {\bibfnamefont {C.~N.}\ \bibnamefont {Lam}}, \bibinfo
  {author} {\bibfnamefont {B.~G.}\ \bibnamefont {Sumpter}}, \ and\ \bibinfo
  {author} {\bibfnamefont {Y.}~\bibnamefont {Wang}},\ }\href@noop {} {\bibfield
   {journal} {\bibinfo  {journal} {ACS Macro Letters}\ }\textbf {\bibinfo
  {volume} {7}},\ \bibinfo {pages} {190} (\bibinfo {year} {2018})}\BibitemShut
  {NoStop}%
\bibitem [{\citenamefont {Kim}\ \emph {et~al.}(2011)\citenamefont {Kim},
  \citenamefont {Stephanou}, \citenamefont {Edwards},\ and\ \citenamefont
  {Khomami}}]{kim2011mean}%
  \BibitemOpen
  \bibfield  {author} {\bibinfo {author} {\bibfnamefont {J.}~\bibnamefont
  {Kim}}, \bibinfo {author} {\bibfnamefont {P.}~\bibnamefont {Stephanou}},
  \bibinfo {author} {\bibfnamefont {B.}~\bibnamefont {Edwards}}, \ and\
  \bibinfo {author} {\bibfnamefont {B.}~\bibnamefont {Khomami}},\ }\href@noop
  {} {\bibfield  {journal} {\bibinfo  {journal} {Journal of Non-Newtonian Fluid
  Mechanics}\ }\textbf {\bibinfo {volume} {166}},\ \bibinfo {pages} {593}
  (\bibinfo {year} {2011})}\BibitemShut {NoStop}%
\bibitem [{\citenamefont {Huang}, \citenamefont {Gompper},\ and\ \citenamefont
  {Winkler}(2012)}]{huang2012non}%
  \BibitemOpen
  \bibfield  {author} {\bibinfo {author} {\bibfnamefont {C.-C.}\ \bibnamefont
  {Huang}}, \bibinfo {author} {\bibfnamefont {G.}~\bibnamefont {Gompper}}, \
  and\ \bibinfo {author} {\bibfnamefont {R.~G.}\ \bibnamefont {Winkler}},\
  }\href@noop {} {\bibfield  {journal} {\bibinfo  {journal} {Journal of
  Physics: Conference Series}\ }\textbf {\bibinfo {volume} {392}},\ \bibinfo
  {pages} {012003} (\bibinfo {year} {2012})}\BibitemShut {NoStop}%
\bibitem [{\citenamefont {Nafar~Sefiddashti}, \citenamefont {Edwards},\ and\
  \citenamefont {Khomami}(2015)}]{nafar2015individual}%
  \BibitemOpen
  \bibfield  {author} {\bibinfo {author} {\bibfnamefont {M.}~\bibnamefont
  {Nafar~Sefiddashti}}, \bibinfo {author} {\bibfnamefont {B.}~\bibnamefont
  {Edwards}}, \ and\ \bibinfo {author} {\bibfnamefont {B.}~\bibnamefont
  {Khomami}},\ }\href@noop {} {\bibfield  {journal} {\bibinfo  {journal}
  {Journal of Rheology}\ }\textbf {\bibinfo {volume} {59}},\ \bibinfo {pages}
  {119} (\bibinfo {year} {2015})}\BibitemShut {NoStop}%
\bibitem [{\citenamefont {Mohagheghi}\ and\ \citenamefont
  {Khomami}(2016)}]{mohagheghi2016elucidating}%
  \BibitemOpen
  \bibfield  {author} {\bibinfo {author} {\bibfnamefont {M.}~\bibnamefont
  {Mohagheghi}}\ and\ \bibinfo {author} {\bibfnamefont {B.}~\bibnamefont
  {Khomami}},\ }\href@noop {} {\bibfield  {journal} {\bibinfo  {journal}
  {Journal of Rheology}\ }\textbf {\bibinfo {volume} {60}},\ \bibinfo {pages}
  {849} (\bibinfo {year} {2016})}\BibitemShut {NoStop}%
\bibitem [{\citenamefont {Sefiddashti}, \citenamefont {Edwards},\ and\
  \citenamefont {Khomami}(2017)}]{sefiddashti2017evaluation}%
  \BibitemOpen
  \bibfield  {author} {\bibinfo {author} {\bibfnamefont {M.~H.~N.}\
  \bibnamefont {Sefiddashti}}, \bibinfo {author} {\bibfnamefont {B.~J.}\
  \bibnamefont {Edwards}}, \ and\ \bibinfo {author} {\bibfnamefont
  {B.}~\bibnamefont {Khomami}},\ }\href@noop {} {\bibfield  {journal} {\bibinfo
   {journal} {Physical Review Fluids}\ }\textbf {\bibinfo {volume} {2}},\
  \bibinfo {pages} {083301} (\bibinfo {year} {2017})}\BibitemShut {NoStop}%
\bibitem [{\citenamefont {Korolkovas}, \citenamefont {Gutfreund},\ and\
  \citenamefont {Barrat}(2016)}]{korolkovas2016simulation}%
  \BibitemOpen
  \bibfield  {author} {\bibinfo {author} {\bibfnamefont {A.}~\bibnamefont
  {Korolkovas}}, \bibinfo {author} {\bibfnamefont {P.}~\bibnamefont
  {Gutfreund}}, \ and\ \bibinfo {author} {\bibfnamefont {J.-L.}\ \bibnamefont
  {Barrat}},\ }\href@noop {} {\bibfield  {journal} {\bibinfo  {journal} {The
  Journal of Chemical Physics}\ }\textbf {\bibinfo {volume} {145}},\ \bibinfo
  {pages} {124113} (\bibinfo {year} {2016})}\BibitemShut {NoStop}%
\bibitem [{\citenamefont {Korolkovas}(2017)}]{airidas_korolkovas_2017_439529}%
  \BibitemOpen
  \bibfield  {author} {\bibinfo {author} {\bibfnamefont {A.}~\bibnamefont
  {Korolkovas}},\ }\href {\doibase 10.5281/zenodo.439529} {\enquote {\bibinfo
  {title} {Simulation of long polymer chains under shear flow},}\ }\bibinfo
  {howpublished} {\url{https://doi.org/10.5281/zenodo.439529}} (\bibinfo {year}
  {2017})\BibitemShut {NoStop}%
\bibitem [{\citenamefont {Bolhuis}\ \emph {et~al.}(2001)\citenamefont
  {Bolhuis}, \citenamefont {Louis}, \citenamefont {Hansen},\ and\ \citenamefont
  {Meijer}}]{bolhuis2001accurate}%
  \BibitemOpen
  \bibfield  {author} {\bibinfo {author} {\bibfnamefont {P.}~\bibnamefont
  {Bolhuis}}, \bibinfo {author} {\bibfnamefont {A.}~\bibnamefont {Louis}},
  \bibinfo {author} {\bibfnamefont {J.}~\bibnamefont {Hansen}}, \ and\ \bibinfo
  {author} {\bibfnamefont {E.}~\bibnamefont {Meijer}},\ }\href@noop {}
  {\bibfield  {journal} {\bibinfo  {journal} {The Journal of Chemical Physics}\
  }\textbf {\bibinfo {volume} {114}},\ \bibinfo {pages} {4296} (\bibinfo {year}
  {2001})}\BibitemShut {NoStop}%
\bibitem [{\citenamefont {de~Gennes}(1979)}]{deGennesScalingConcepts}%
  \BibitemOpen
  \bibfield  {author} {\bibinfo {author} {\bibfnamefont {P.~G.}\ \bibnamefont
  {de~Gennes}},\ }\href@noop {} {\emph {\bibinfo {title} {Scaling Concepts in
  Polymer Physics}}}\ (\bibinfo  {publisher} {Cornell University Press},\
  \bibinfo {year} {1979})\BibitemShut {NoStop}%
\bibitem [{\citenamefont {Sun}\ and\ \citenamefont
  {Faller}(2005)}]{sun2005systematic}%
  \BibitemOpen
  \bibfield  {author} {\bibinfo {author} {\bibfnamefont {Q.}~\bibnamefont
  {Sun}}\ and\ \bibinfo {author} {\bibfnamefont {R.}~\bibnamefont {Faller}},\
  }\href@noop {} {\bibfield  {journal} {\bibinfo  {journal} {Computers \&
  chemical engineering}\ }\textbf {\bibinfo {volume} {29}},\ \bibinfo {pages}
  {2380} (\bibinfo {year} {2005})}\BibitemShut {NoStop}%
\bibitem [{\citenamefont {Kremer}\ and\ \citenamefont
  {Grest}(1990)}]{kremer1990dynamics}%
  \BibitemOpen
  \bibfield  {author} {\bibinfo {author} {\bibfnamefont {K.}~\bibnamefont
  {Kremer}}\ and\ \bibinfo {author} {\bibfnamefont {G.~S.}\ \bibnamefont
  {Grest}},\ }\href@noop {} {\bibfield  {journal} {\bibinfo  {journal} {The
  Journal of Chemical Physics}\ }\textbf {\bibinfo {volume} {92}},\ \bibinfo
  {pages} {5057} (\bibinfo {year} {1990})}\BibitemShut {NoStop}%
\bibitem [{\citenamefont {Cao}\ and\ \citenamefont
  {Likhtman}(2012)}]{cao2012shear}%
  \BibitemOpen
  \bibfield  {author} {\bibinfo {author} {\bibfnamefont {J.}~\bibnamefont
  {Cao}}\ and\ \bibinfo {author} {\bibfnamefont {A.~E.}\ \bibnamefont
  {Likhtman}},\ }\href@noop {} {\bibfield  {journal} {\bibinfo  {journal}
  {Physical review letters}\ }\textbf {\bibinfo {volume} {108}},\ \bibinfo
  {pages} {028302} (\bibinfo {year} {2012})}\BibitemShut {NoStop}%
\bibitem [{\citenamefont {Ba{\v c}ov{\'a}}\ \emph {et~al.}(2013)\citenamefont
  {Ba{\v c}ov{\'a}}, \citenamefont {Hawke}, \citenamefont {Read},\ and\
  \citenamefont {Moreno}}]{bacova2013dynamics}%
  \BibitemOpen
  \bibfield  {author} {\bibinfo {author} {\bibfnamefont {P.}~\bibnamefont
  {Ba{\v c}ov{\'a}}}, \bibinfo {author} {\bibfnamefont {L.~G.}\ \bibnamefont
  {Hawke}}, \bibinfo {author} {\bibfnamefont {D.~J.}\ \bibnamefont {Read}}, \
  and\ \bibinfo {author} {\bibfnamefont {A.~J.}\ \bibnamefont {Moreno}},\
  }\href@noop {} {\bibfield  {journal} {\bibinfo  {journal} {Macromolecules}\
  }\textbf {\bibinfo {volume} {46}},\ \bibinfo {pages} {4633} (\bibinfo {year}
  {2013})}\BibitemShut {NoStop}%
\bibitem [{\citenamefont {Ilg}, \citenamefont {{\"O}ttinger},\ and\
  \citenamefont {Kr{\"o}ger}(2009)}]{ilg2009systematic}%
  \BibitemOpen
  \bibfield  {author} {\bibinfo {author} {\bibfnamefont {P.}~\bibnamefont
  {Ilg}}, \bibinfo {author} {\bibfnamefont {H.~C.}\ \bibnamefont
  {{\"O}ttinger}}, \ and\ \bibinfo {author} {\bibfnamefont {M.}~\bibnamefont
  {Kr{\"o}ger}},\ }\href@noop {} {\bibfield  {journal} {\bibinfo  {journal}
  {Physical Review E}\ }\textbf {\bibinfo {volume} {79}},\ \bibinfo {pages}
  {011802} (\bibinfo {year} {2009})}\BibitemShut {NoStop}%
\bibitem [{\citenamefont {Ramesh}(2008)}]{ramesh2008}%
  \BibitemOpen
  \bibfield  {author} {\bibinfo {author} {\bibfnamefont {K.}~\bibnamefont
  {Ramesh}},\ }in\ \href@noop {} {\emph {\bibinfo {booktitle} {Springer
  handbook of experimental solid mechanics}}},\ \bibinfo {editor} {edited by\
  \bibinfo {editor} {\bibfnamefont {W.~N.}\ \bibnamefont {Sharpe}}}\ (\bibinfo
  {publisher} {Springer Science \& Business Media},\ \bibinfo {address} {New
  York},\ \bibinfo {year} {2008})\ Chap.~\bibinfo {chapter} {33}, pp.\ \bibinfo
  {pages} {929--960}\BibitemShut {NoStop}%
\bibitem [{\citenamefont {Nghe}, \citenamefont {Tabeling},\ and\ \citenamefont
  {Ajdari}(2010)}]{nghe2010flow}%
  \BibitemOpen
  \bibfield  {author} {\bibinfo {author} {\bibfnamefont {P.}~\bibnamefont
  {Nghe}}, \bibinfo {author} {\bibfnamefont {P.}~\bibnamefont {Tabeling}}, \
  and\ \bibinfo {author} {\bibfnamefont {A.}~\bibnamefont {Ajdari}},\
  }\href@noop {} {\bibfield  {journal} {\bibinfo  {journal} {Journal of
  Non-Newtonian Fluid Mechanics}\ }\textbf {\bibinfo {volume} {165}},\ \bibinfo
  {pages} {313} (\bibinfo {year} {2010})}\BibitemShut {NoStop}%
\bibitem [{\citenamefont {Padding}\ and\ \citenamefont
  {Briels}(2001)}]{padding2001uncrossability}%
  \BibitemOpen
  \bibfield  {author} {\bibinfo {author} {\bibfnamefont {J.}~\bibnamefont
  {Padding}}\ and\ \bibinfo {author} {\bibfnamefont {W.~J.}\ \bibnamefont
  {Briels}},\ }\href@noop {} {\bibfield  {journal} {\bibinfo  {journal} {The
  Journal of Chemical Physics}\ }\textbf {\bibinfo {volume} {115}},\ \bibinfo
  {pages} {2846} (\bibinfo {year} {2001})}\BibitemShut {NoStop}%
\bibitem [{\citenamefont {Karayiannis}\ and\ \citenamefont
  {Kr{\"o}ger}(2009)}]{karayiannis2009combined}%
  \BibitemOpen
  \bibfield  {author} {\bibinfo {author} {\bibfnamefont {N.~C.}\ \bibnamefont
  {Karayiannis}}\ and\ \bibinfo {author} {\bibfnamefont {M.}~\bibnamefont
  {Kr{\"o}ger}},\ }\href@noop {} {\bibfield  {journal} {\bibinfo  {journal}
  {International journal of molecular sciences}\ }\textbf {\bibinfo {volume}
  {10}},\ \bibinfo {pages} {5054} (\bibinfo {year} {2009})}\BibitemShut
  {NoStop}%
\bibitem [{\citenamefont {Kawecki}\ \emph {et~al.}(2016)\citenamefont
  {Kawecki}, \citenamefont {Gutfreund}, \citenamefont {Adlmann}, \citenamefont
  {Lindholm}, \citenamefont {Longeville}, \citenamefont {Lapp},\ and\
  \citenamefont {Wolff}}]{kawecki2016probing}%
  \BibitemOpen
  \bibfield  {author} {\bibinfo {author} {\bibfnamefont {M.}~\bibnamefont
  {Kawecki}}, \bibinfo {author} {\bibfnamefont {P.}~\bibnamefont {Gutfreund}},
  \bibinfo {author} {\bibfnamefont {F.}~\bibnamefont {Adlmann}}, \bibinfo
  {author} {\bibfnamefont {E.}~\bibnamefont {Lindholm}}, \bibinfo {author}
  {\bibfnamefont {S.}~\bibnamefont {Longeville}}, \bibinfo {author}
  {\bibfnamefont {A.}~\bibnamefont {Lapp}}, \ and\ \bibinfo {author}
  {\bibfnamefont {M.}~\bibnamefont {Wolff}},\ }\href@noop {} {\bibfield
  {journal} {\bibinfo  {journal} {Journal of Physics: Conference Series}\
  }\textbf {\bibinfo {volume} {746}},\ \bibinfo {pages} {012014} (\bibinfo
  {year} {2016})}\BibitemShut {NoStop}%
\bibitem [{\citenamefont {Hsu}\ and\ \citenamefont
  {Kremer}(2017)}]{hsu2017detailed}%
  \BibitemOpen
  \bibfield  {author} {\bibinfo {author} {\bibfnamefont {H.-P.}\ \bibnamefont
  {Hsu}}\ and\ \bibinfo {author} {\bibfnamefont {K.}~\bibnamefont {Kremer}},\
  }\href@noop {} {\bibfield  {journal} {\bibinfo  {journal} {The European
  Physical Journal Special Topics}\ }\textbf {\bibinfo {volume} {226}},\
  \bibinfo {pages} {693} (\bibinfo {year} {2017})}\BibitemShut {NoStop}%
\bibitem [{\citenamefont {Unidad}\ and\ \citenamefont
  {Ianniruberto}(2014)}]{Unidad2014}%
  \BibitemOpen
  \bibfield  {author} {\bibinfo {author} {\bibfnamefont {H.~J.}\ \bibnamefont
  {Unidad}}\ and\ \bibinfo {author} {\bibfnamefont {G.}~\bibnamefont
  {Ianniruberto}},\ }\href {\doibase 10.1007/s00397-013-0755-x} {\bibfield
  {journal} {\bibinfo  {journal} {Rheologica Acta}\ }\textbf {\bibinfo {volume}
  {53}},\ \bibinfo {pages} {191} (\bibinfo {year} {2014})}\BibitemShut
  {NoStop}%
\bibitem [{\citenamefont {Cleyn}\ and\ \citenamefont
  {Mewis}(1981)}]{Cleyn1981}%
  \BibitemOpen
  \bibfield  {author} {\bibinfo {author} {\bibfnamefont {G.~D.}\ \bibnamefont
  {Cleyn}}\ and\ \bibinfo {author} {\bibfnamefont {J.}~\bibnamefont {Mewis}},\
  }\href {\doibase https://doi.org/10.1016/0377-0257(87)87009-X} {\bibfield
  {journal} {\bibinfo  {journal} {Journal of Non-Newtonian Fluid Mechanics}\
  }\textbf {\bibinfo {volume} {9}},\ \bibinfo {pages} {91 } (\bibinfo {year}
  {1981})}\BibitemShut {NoStop}%
\bibitem [{\citenamefont {Kim}\ \emph {et~al.}(2013)\citenamefont {Kim},
  \citenamefont {Mewis}, \citenamefont {Clasen},\ and\ \citenamefont
  {Vermant}}]{kim2013superposition}%
  \BibitemOpen
  \bibfield  {author} {\bibinfo {author} {\bibfnamefont {S.}~\bibnamefont
  {Kim}}, \bibinfo {author} {\bibfnamefont {J.}~\bibnamefont {Mewis}}, \bibinfo
  {author} {\bibfnamefont {C.}~\bibnamefont {Clasen}}, \ and\ \bibinfo {author}
  {\bibfnamefont {J.}~\bibnamefont {Vermant}},\ }\href@noop {} {\bibfield
  {journal} {\bibinfo  {journal} {Rheologica Acta}\ }\textbf {\bibinfo {volume}
  {52}},\ \bibinfo {pages} {727} (\bibinfo {year} {2013})}\BibitemShut
  {NoStop}%
\bibitem [{\citenamefont {Harasim}\ \emph {et~al.}(2013)\citenamefont
  {Harasim}, \citenamefont {Wunderlich}, \citenamefont {Peleg}, \citenamefont
  {Kr{\"o}ger},\ and\ \citenamefont {Bausch}}]{harasim2013direct}%
  \BibitemOpen
  \bibfield  {author} {\bibinfo {author} {\bibfnamefont {M.}~\bibnamefont
  {Harasim}}, \bibinfo {author} {\bibfnamefont {B.}~\bibnamefont {Wunderlich}},
  \bibinfo {author} {\bibfnamefont {O.}~\bibnamefont {Peleg}}, \bibinfo
  {author} {\bibfnamefont {M.}~\bibnamefont {Kr{\"o}ger}}, \ and\ \bibinfo
  {author} {\bibfnamefont {A.~R.}\ \bibnamefont {Bausch}},\ }\href@noop {}
  {\bibfield  {journal} {\bibinfo  {journal} {Physical review letters}\
  }\textbf {\bibinfo {volume} {110}},\ \bibinfo {pages} {108302} (\bibinfo
  {year} {2013})}\BibitemShut {NoStop}%
\bibitem [{\citenamefont {Kobayashi}\ and\ \citenamefont
  {Yamamoto}(2011)}]{kobayashi2011implementation}%
  \BibitemOpen
  \bibfield  {author} {\bibinfo {author} {\bibfnamefont {H.}~\bibnamefont
  {Kobayashi}}\ and\ \bibinfo {author} {\bibfnamefont {R.}~\bibnamefont
  {Yamamoto}},\ }\href@noop {} {\bibfield  {journal} {\bibinfo  {journal} {The
  Journal of chemical physics}\ }\textbf {\bibinfo {volume} {134}},\ \bibinfo
  {pages} {064110} (\bibinfo {year} {2011})}\BibitemShut {NoStop}%
\end{thebibliography}%

\end{document}